\newif \ifdraft \draftfalse
\newif \iffull \fullfalse
\newif \ifanonymous \anonymousfalse

\documentclass[nonacm,acmsmall,\ifanonymous,review,anonymous\fi]{acmart}

\usepackage[T1]{fontenc}

\acmJournal{PACMPL}
\acmVolume{1}
\acmNumber{CONF} 
\acmArticle{1}
\acmYear{2018}
\acmMonth{1}
\acmDOI{} 
\startPage{1}

\setcopyright{none}

\usepackage{booktabs}
\usepackage{enumitem}
\usepackage{subcaption}
\usepackage{amsmath,amssymb}
\usepackage{listings}
\usepackage{lstocaml}
\usepackage{lstsh}
\usepackage{bcprules}
\usepackage{tikz}
\usetikzlibrary{arrows,shapes,positioning,chains}
\usepackage{fp}
\usepackage{local}

\makeatletter
\newcommand{\graf}{\@startsection{paragraph}{4}{0\p@}
  {-.25\baselineskip}
  {-3.5\p@}%
  {\ACM@NRadjust{\@parfont\@adddotafter}}}
\makeatother

\newcommand{\YES}{$\checkmark$}
\newcommand{\NO}{$\times$}

\newcommand{\fullbreak}{\iffull \\ \fi}

\newcommand{\FS}{\texttt{\textvisiblespace}\xspace}
\renewcommand{\tilde}{\texttt{\textasciitilde}\xspace}
\renewcommand{\star}{\texttt{*}\xspace}
\newcommand{\at}{\texttt{@}\xspace}
\newcommand{\amp}{\texttt{\&}\xspace}
\newcommand{\param}[2]{\texttt{\textdollar\{\ensuremath{#1} ~ \ensuremath{#2}\}}\xspace}
\newcommand{\backtick}[1]{\texttt{\textdollar(\ensuremath{#1})}\xspace}
\newcommand{\arith}[1]{\texttt{\textdollar((\ensuremath{#1}))}\xspace}
\renewcommand{\quote}[1]{\texttt{"\ensuremath{#1}"}\xspace}
\newcommand{\exitstatus}{\texttt{\textdollar{}?}}

\newcommand{\lambdajs}{$\lambda_{\mathrm{JS}}$\xspace}
\newcommand{\dash}{\texttt{dash}\xspace}
\newcommand{\bash}{\texttt{bash}\xspace}
\newcommand{\yash}{\texttt{yash}\xspace}
\newcommand{\zsh}{\texttt{zsh}\xspace}
\newcommand{\ksh}{\texttt{ksh}\xspace}
\newcommand{\mksh}{\texttt{mksh}\xspace}
\newcommand{\OSH}{\texttt{OSH}\xspace}

\newcommand{\fd}{\texttt{fd}\xspace}
\newcommand{\pid}{\texttt{pid}\xspace}
\newcommand{\sig}{\texttt{sig}\xspace}

\newcommand{\bcheck}{\ensuremath{b_{\checkmark}}\xspace}
\newcommand{\subst}{\ensuremath{\texttt{\textdollar{}()}}\xspace}
\newcommand{\bsubst}{\ensuremath{b_{\subst}}\xspace}

\newcommand{\helper}[1]{\ensuremath{\mathsf{#1}}\xspace}
\newcommand{\syscall}[1]{\texttt{#1}\xspace}

\newcommand{\expError}{\helper{expError}}
\newcommand{\setParam}{\helper{setParam}}
\newcommand{\setLocal}{\helper{setLocal}}
\newcommand{\toString}{\helper{toString}}
\newcommand{\checkTraps}{\helper{checkTraps}}

\newcommand{\blindurl}[1]{\footnote{
\ifanonymous
URL suppressed for double blind review.
\else
\url{#1}
\fi
}}

\begin{document}

\title[Smoosh: executable formal semantics of the POSIX shell]{Executable formal semantics for the POSIX shell}
\subtitle{Smoosh: the Symbolic, Mechanized, Observable, Operational Shell}         


\author{Michael Greenberg}
\affiliation{
  \department{Department of Computer Science}              
  \institution{Pomona College}            
  \city{Claremont}
  \state{CA}
  \country{USA}                    
}
\email{michael@cs.pomona.edu}          

\author{Austin J. Blatt}
\authornote{Work done while at Pomona College.}          
\affiliation{
  \institution{Puppet Labs}           
  \city{Portland}
  \state{OR}
  \country{USA}                   
}
\email{austinblatt@gmail.com}         

\begin{abstract}
The POSIX shell is a widely deployed, powerful tool for managing
computer systems. The shell is the expert's control panel, a
necessary tool for configuring, compiling, installing, maintaining,
and deploying systems.
Even though it is powerful, critical infrastructure, the POSIX shell is
maligned and misunderstood. Its power and its subtlety are a dangerous
combination.

We define a formal, mechanized, executable small-step semantics for
the POSIX shell, which we call Smoosh.
We compared Smoosh against seven other shells that aim for some
measure of POSIX compliance (\bash, \dash, \zsh, \OSH, \mksh, \ksh{}93,
and \yash). Using three test suites---the POSIX test suite, the
Modernish test suite and shell diagnosis, and a test suite of our own
device---we found
Smoosh's semantics to be the most conformant to the
POSIX standard. Modernish judges Smoosh to have the fewest bugs (just
one, from using \dash's parser) and no quirks.
To show that our semantics is useful beyond yielding a conformant,
executable shell, we also implemented a symbolic stepper to
illuminate the subtle behavior of the shell.

Smoosh will serve as a foundation for formal study of the POSIX shell,
supporting research on and development of new shells, new tooling for
shells, and new shell designs.
\end{abstract}

%



\maketitle

\section{Introduction}
\label{sec:intro}

The POSIX shell is a line-oriented, potentially interactive scripting
language~\cite{POSIXbase}.
The shell is widely used by experts on all three major platforms
(Linux, OS X, and Windows);
it is often the best---or only!---way to configure, compile, install,
manage, deploy, or remove software systems.
The shell is used for these tasks because the shell easily combines
filesystem manipulations (using standard utilities like \texttt{cp}),
system management features (like \texttt{apt}), and asynchronous job
control (using pipelines $c_1 ~|~ c_2 ~|~ \dots$, background jobs $c ~
\amp$, and the \texttt{wait} builtin).
Modern container systems rely on the shell: some use the shell
extensively (e.g., Docker) while others use the shell as a necessary
escape hatch (e.g., Vagrant, Puppet).
The POSIX shell is critical software infrastructure.

Critical though it may be, the POSIX shell is not well understood.
It has unusual semantics: the general command language doesn't
evaluate subexpressions, but rather treats them as strings and
\emph{expands} the special control codes in them.
This process, called \emph{word expansion}, is responsible for: 
the translation of $\tilde$ into one's home directory; 
parameter, i.e., variable substitution; 
command substitution (written $\backtick{c}$ or
$\texttt{`}c\texttt{`}$); 
and globbing with \star and \texttt{?}.
Word expansion is simultaneously part of the shell's power and
part of its danger~\cite{Greenberg18px}.
It is very easy for word expansion to generate too many or too few
arguments to a command... and commands play for keeps! One famous
example is ``Steam
cleaning'',\footnote{A misunderstanding of expansion in a Steam script wiped hard drives: \url{http://www.theregister.co.uk/2015/01/17/scary_code_of_the_week_steam_cleans_linux_pcs/}.}
though similar bugs abound.
The treacherous combination of powerful commands and abstruse semantics
has led to no small amount of disgust with the shell (see, for
example, the \textit{UNIX Hater's Handbook}~\cite{UHH}).

\medskip

\noindent
%
We aim to rehabilitate the shell. The POSIX shell is one of the
oldest programming languages still in popular use; it ought to have
tool support commensurate with the state of the art.
In order to have effective tool support, we need to be able to soundly reason about the shell's
semantics: to summarize scripts' behavior; to
calculate scripts' dependencies or preconditions; or to compile
scripts to be faster, safer, or in another language.

To that end, we introduce Smoosh: the \underline{s}ymbolic, \underline{m}echanized, \underline{o}bservable, \underline{o}perational \underline{sh}ell.
Smoosh serves as a \emph{mechanized} reference semantics for the POSIX standard.
Smoosh is, first and foremost, an \emph{operational} shell: it can be used interactively or for scripting.
Smoosh is also \emph{symbolic} and \emph{observable}, generating
traces given in real and simulated environments.

Even though it is a completely usable shell, the mechanized small-step
operational semantics for Smoosh is relatively compact: the two core
stepping functions (Section~\ref{sec:impl}) are a combined total of
1\,034 SLOC of Lem~\cite{Mulligan:2014:LRE:2628136.2628143}.
The small-step nature of the semantics makes it easy to capture and
report information. Smoosh can, e.g., detect and report unspecified
and undefined behaviors, trace how and when system calls are made, and
log signal handling and traps.

Executability is critical in order to validate our semantics by \emph{testing}.
Smoosh passes all of the locale-independent parts of the POSIX test suite.
We compared Smoosh against seven other shells that aim for some
measure of POSIX compliance (\bash, \dash, \zsh, \OSH, \mksh, \ksh{}93,
and \yash). We used three test suites: the POSIX test suite, the
Modernish test suite and shell diagnosis~\cite{Modernish}, and a test
suite of our own device.
On all three suites, Smoosh's semantics is the most conformant to the
POSIX standard; Modernish judges Smoosh to be the
least buggy or quirky. (See Section~\ref{sec:results} for more detail.)

Smoosh isn't just an executable shell, though: Smoosh is parameterized
over the operating system state, highlighting the OS interface
demanded by a POSIX shell and allowing for alternative uses of the
semantics.
We've implemented two possible choices of the OS state: a
\texttt{system} mode where system calls actually occur,
and a \texttt{symbolic} mode where the system calls are
simulated.
We use \texttt{symbolic} mode to implement a stepper for the shell
(Section~\ref{sec:stepper}).

\smallskip
\noindent
We claim the following contributions:
\begin{itemize}
\item Smoosh is a new implementation of the POSIX shell specification (Section~\ref{sec:what}). Its
  semantics is faithful while still being of manageable size
  (Section~\ref{sec:semantics}).
\item We parameterize the Smoosh semantics over a suite of operating
  system functions (Section~\ref{sec:impl}). The parameter is
  theoretically useful as it creates a barrier between the shell and
  the OS; it is practically useful in that it allows for alternative
  instantiations of OS primitives. As an example, we instantiate the
  OS with a symbolic mode to write a program stepper, which
  illuminates the shell's obscure word expansion and evaluation semantics
  (Section~\ref{sec:stepper}).
\item Smoosh is conformant to the POSIX standard and suitable
  as a canonical implementation. 
  Smoosh is the most conformant and least buggy of the seven shells tested (Section~\ref{sec:results},
  Table~\ref{tab:tests}).
\item The development of Smoosh led to the identification of numerous
  bugs in shells in active use (\bash, \dash, \yash, \OSH) and
  in the POSIX specification and its test suite (Section~\ref{sec:bugs}).
\end{itemize}
These contributions lay the foundation for serious tools for improving
the shell. Without a semantics, what could we prove about an
analysis or compiler for the shell? Smoosh's semantics 
it 
can be a reference implementation of POSIX for desugarers (like
CoLiS~\cite{jeannerod:hal-01432034}) and compilers to test and prove
against,
as well as a semantics against which to prove static analyses sound.
Our tested semantics will be
a baseline for work improving on the shell's implementation
and design. 

\section{What is the POSIX shell?}
\label{sec:what}

There are two perspectives on what defines the POSIX shell: one places the standard first, and the other places shell implementations first. We aim for Smoosh to accommodate both perspectives.

The first, bureaucratic perspective centers the POSIX spec (\cite{POSIXbase}, Volume ``Shell \& Utilities''), which
comprises 21 pages of introduction (\S{}1), 98 pages of core
definitions (\S{}2 ``Shell Command Language''), and 
additional documentation on 160 utilities (ranging in scale from
\texttt{awk} to \texttt{true}).\footnote{We write out `Section' when
  referring to sections in this paper and use \S{} to refer to sections
  of the POSIX specification.}
The part most relevant to shell implementors and programmers is
Volume ``Shell \& Utilities'' \S{}2 ``Shell Command Language'',
which weaves together explanation of the POSIX shell's lexer,
parser, and semantics.
Some important information is left to the description of the
\texttt{sh} command in \S{}4.
The spec is written in the typical style, with specific definitions of
words like `can'\iffull (``the feature or behavior is mandatory for an
implementation that conforms to POSIX.1-2017'')\fi, `may'\iffull (``a feature or
behavior that is optional for an implementation that conforms to
POSIX.1-2017'')\fi, and that troublesome pair, `unspecified' (``a value
or behavior not specified by POSIX.1-2017 which results from use of a
valid program construct or valid data input'') and `undefined' (``\iffull{}a
value or behavior not defined by POSIX.1-2017 \else \dots \fi which results from use
of an invalid program construct or invalid data
input'').\iffull\footnote{\S{}3 ``Batch Environment Services'' is
  obsolescent as of Issue 7 (the current version of the POSIX spec).}\fi{}
There are 70 uses of `unspecified' and 17 uses of `undefined' relevant to
the shell.

The second, pragmatic perspective is that the POSIX shell is a
collection of language implementations (e.g., \bash, \dash, \yash)
that more or less agree on a core set of features; the POSIX
specification is a document that tries to track what that core set is,
knowing that few implementations will conform in all respects.

We consider both of these points of view when we discuss conformance
(Section~\ref{sec:results}).
When we refer to shell implementations, we mean the versions in Debian
9 stable: \bash 4.4-12(1), \dash 0.5.6-2.4, \zsh 5.3.1-4+b2, \OSH
0.6.pre21, \mksh 54-2+b4, \ksh{}93 20120801-3.1, and \yash 2.43-1.

There is already precedent for taking something like
the POSIX spec and producing a formal model, with recent work on C
being a good
example~\cite{Blazy2009,Ellison:2012:EFS:2103656.2103719,10.1007/978-3-319-08970-6_36,Kang:2015:FCM:2737924.2738005,Memarian:2016:DCE:2908080.2908081}.
There is similar precedent for taking a collection of implementations and producing a formal model, with \lambdajs being a good example~\cite{10.1007/978-3-642-14107-2_7}.
What is special about doing this process for the POSIX shell?
The shell has two distinctive features that make its semantics more
interesting:
\begin{itemize}[wide, leftmargin=*, labelindent=0pt]
\item \textbf{Word expansion.} While typical languages evaluate an
  expression by evaluating its subexpressions, many shell expressions
  evaluate their parts by performing \emph{word expansion}
  rather than recursive evaluation. Many familiar shell features are
  implemented via expansion: \tilde expanding to your home directory;
  variables \verb|$x| expanding to their contents; command
  substitutions \backtick{c} expanding to the captured \texttt{STDOUT}
  of the inner command. (See Section~\ref{sec:exp-informal}.)
\item \textbf{System calls.} While typical languages make system calls
  via library functions, the shell makes system calls in its
  semantics. Core operations depend on forking a new
  process (\texttt{fork}), replacing the current process with an
  executable (\texttt{execve}), waiting on processes (\texttt{wait}
  and \texttt{waitpid}), handling and sending signals
  (\texttt{signal}, \texttt{kill}), working with files (\texttt{open},
  \texttt{close}, \texttt{access}, \texttt{stat}, \texttt{lstat},
  \texttt{write}, \texttt{read}), and creating and manipulating file
  descriptors (\texttt{pipe}, \texttt{fcntl}, \texttt{dup2}).
\end{itemize}
These two features present different problems when trying to
understand the shell well enough to construct a sound working model.
Modeling expansion is tricky because it (a) is mutually recursive with
evaluation; (b) is a subtly structured, four-stage process; and (c)
is used by the evaluation semantics in several slightly different
configurations.
Modeling system calls is tricky because (a) there are many of them, (b) their
behavior is subtle, and (c) giving faithful meaning to them can bring
arbitrarily much of the operating system into the formal model.

\subsection{What is word expansion?}
\label{sec:exp-informal}

\begin{figure}

\begin{subfigure}[b]{.2\linewidth}
\begin{verbatim}
$ echo ~root
/var/root
$ usr=root
$ echo ~$usr
~root
\end{verbatim}
\caption{Tilde expansion}
\label{expeg:tilde}
\end{subfigure}
~~
\begin{subfigure}[b]{.5\linewidth}
\begin{verbatim}
$ x=$(ls)
$ echo $x,${#x},${x#*[ab]},${x##*[ab]}.
a b c,5, b c, c.
\end{verbatim}
\caption{Command substitution, parameter expansion}
\label{expeg:command-param}
\end{subfigure}
~~
\begin{subfigure}[b]{.25\linewidth}
\begin{verbatim}
$ y=42 x=5
$ echo $((y += $x))
47
$ echo $((y)) $y
47 47  
\end{verbatim}
\caption{Arithmetic expansion}
\label{expeg:arithmetic}
\end{subfigure}

~ \\[.4em]

\begin{subfigure}[b]{.4\linewidth}
\begin{verbatim}
$ ls
a b c
$ x="a b"
$ ls $x
a b
$ ls "$x"
ls: a b: No such file or directory
\end{verbatim}
\caption{Field splitting}
\label{expeg:field-splitting}
\end{subfigure}
\quad
\begin{subfigure}[b]{.55\linewidth}
\begin{verbatim}
$ echo a*
ap app appall apparition appendix applejack
$ echo ap?
app
$ echo appa*
appall apparition
$ echo ap[=p=]*a*
appall apparition applejack
$ echo "a*"
a*
\end{verbatim}
\caption{Pathname expansion (a/k/a globbing), quote removal}
\label{expeg:globbing}
\end{subfigure}

  \caption{Expansion examples}
  \label{fig:expansion-examples}
\end{figure}

The POSIX spec indicates that there are four stages of expansion with a total of seven components:
(1) tilde expansion, 
parameter expansion, 
command substitution, 
and arithmetic expansion;
(2) field splitting;
(3) pathname expansion; and 
(4) quote removal.
Shells perform this word expansion left-to-right in stage order (see
Section~\ref{sec:expansion} for Smoosh's semantics).
These seven components are best illustrated by brief examples
(Figure~\ref{fig:expansion-examples}).
For a more detailed description of these phases, we refer readers to
Greenberg's arguments surrounding word expansion~\citeyearpar{Greenberg18px}
and to the POSIX specification~\cite{POSIXbase} \S{}2.6.
The examples in Figures~\ref{expeg:command-param}
and~\ref{expeg:field-splitting} take place in a directory with three
files (\texttt{a}, \texttt{b}, and \texttt{c}); the example in
Figure~\ref{expeg:globbing} takes place in a different directory,
where the result of the first \texttt{echo} indicates which files
begin with the letter \texttt{a}.

\iffull
In \emph{tilde expansion} (Figure~\ref{expeg:tilde}), the string
\tilde is replaced by the current user's directory as indicated by the
\texttt{HOME} variable; one can write \texttt{\tilde{}FOO}, in which
case the shell will use the \texttt{getpwnam} system call to look up
the initial working directory for user \texttt{FOO}.
\iffull
A variety of rules govern when tilde expansion can happen and when a
user is used and when \texttt{HOME} is used.
Normally, tildes are only expanded when
they are the first thing in the word; when in an assignment, however,
tildes are expanded at the beginning and after unquoted colons.
\fi
There are subtle rules surrounding what constitutes a username
following a tilde, the so-called ``tilde prefix''.
Even though tilde expansion cannot be indirect
(Figure~\ref{expeg:tilde}, second \texttt{echo}), some shells
determine the tilde prefix dynamically.
\iffull
We parse it statically, simplifying our presentation of expansion at
no cost to correctness or conformance.
We record the tilde prefix in the AST as the possibly empty string
$s^?$.
Some shells (e.g., \zsh) support this indirection as a non-conformant
extension of POSIX.
\fi

\emph{Parameter expansion} is how the POSIX shell implements
variables. In the simple case, one simply writes
\texttt{\textdollar{}x} or \texttt{\textdollar\{x\}}, which expands
either to the empty string or the current value of \texttt{x}.
Variable names are statically specified, i.e., one cannot write
\texttt{\textdollar{}\{\textdollar{}y\}} to indirectly use \texttt{y}
to lookup a variable.
Some variables are special, particularly \texttt{\textdollar{}*} and
\texttt{\textdollar{}\at} which represent the current positional
parameters, i.e., command-line arguments or function arguments.
The shell also supports \emph{parameter formats}.
\iffull
As a first example,
defaulting, written \texttt{\textdollar\{x-w\}}, will expand to the
contents of the variable \texttt{x} when \texttt{x} is set... but when
it is not, it will expand \texttt{w} instead. 
The related parameter format \texttt{\textdollar\{x:-w\}} will use the
default words \texttt{w} not only when the variable \texttt{x} is
unset, but also when \texttt{x} is set to the null string.
The length format, \texttt{\textdollar\{\#x\}} will expand to the
length of the contents of \texttt{x} (with existing shells universally
choosing to return 0 when \texttt{x} is unset). Unlike other parameter
formats, here the formatter \texttt{\#} comes before the variable
name in the shell's concrete syntax.
\fi
Each shell offers some fixed set of parameter formats extending the
POSIX set of \texttt{-w} for defaulting, \texttt{=w} for defaulting
assignment, \texttt{?w} for erroring, \texttt{+w} for providing an
alternative, \texttt{\#x} for length, \texttt{\#} and \texttt{\#\#} for
smallest and largest prefix removal, and \texttt{\%} and \texttt{\%\%}
for smallest and largest suffix removal.
Figure~\ref{expeg:command-param} shows four: `normal' formatting
(\verb|$x|), the length formatter (\verb|${#x}|), smallest prefix
removal (\verb|${x#*[ab]|), and largest prefix removal
(\verb|${x##*[ab]|).

The prefix and suffix removal parameter formats use \emph{patterns}:
the symbols \texttt{*} and \texttt{?} are treated specially (any
number of characters and a single character, respectively); one can
use brackets specify sets or ranges of characters, either explicitly
as in \texttt{[a-z]} or as classes as in \texttt{[:space:]} for the
any space characters, \texttt{[=e=]} for the letters equivalent to `e'
in the current locale (which may include `\'e' in, say, French), or
\texttt{[.ll.]} for the set of letters that sort with the digraph `ll'
(which is treated separately in Spanish).

\emph{Command substitution} is how shell scripts can evaluate
subexpressions rather than merely expanding them. Written \texttt{`cmd
  arg1 ...`} or \backtick{\texttt{cmd arg1 ...}}, command substitution works by
running \texttt{cmd arg1 ...} and capturing the output on STDOUT; the
captured output is spliced back in to the result command.
Figure~\ref{expeg:command-param} uses command
substitution to set \texttt{x} to the output of \texttt{ls}.

\emph{Arithmetic expansion} allows shell programmers to manipulate
twos-complement signed integers. When handling an arithmetic
expression \arith{e}, the expression \texttt{e} is fully expanded
before being parsed as a C-style arithmetic expression.
Arithmetic expressions include mutations, like \texttt{x = e},
\texttt{x += e}, and \texttt{x++} (Figure~\ref{expeg:arithmetic}).
Perhaps confusingly, variables may or may not be written with a \texttt{\textdollar}. Since the arithmetic expression is fully expanded before being parsed, variables written with \texttt{\textdollar} will be expanded by parameter expansion and their resulting values will be parsed; variables without \texttt{\textdollar} will be parsed as variables and evaluated later.
\iffull
For example, \arith{z \mathrel{\texttt{+=}} \texttt{\textdollar{}}x * y} will, when \texttt{x} is set to 5, first expand to \texttt{z += 5 * y} and later consult the values of \texttt{y} and \texttt{z} when evaluating the arithmetic expression.
Unset and null variables are treated as 0; variables set to
non-numeric values cause an error.
\fi

\emph{Field splitting} is the process by which the results of various expansions are broken up into constituent fields for commands or \texttt{for} loops.
The default is break on spaces, tabs, and newlines. Quotation with double quotes prevents field splitting. Without quotes, \texttt{\textdollar{}x} expands to two fields (Figure~\ref{expeg:field-splitting}), and so
\texttt{ls} gets two arguments: \texttt{a} and \texttt{b}. With
quotes, field splitting never happens, and so \texttt{ls} gets just
one argument: \texttt{a\FS{}b}, where \FS is a space.
The characters uses to split up fields can be controlled with the \texttt{IFS} variable. When
unset, \texttt{IFS} defaults to space, newline, and tab.
\iffull Setting \texttt{IFS=,} allows for processing of (simple,
unquoted) CSV; setting \texttt{IFS=:} allows for processing of colon
separated paths, as in the \texttt{PATH} variable.\fi

\emph{Pathname expansion} is better known by its colloquial name of
\emph{globbing}. Globbing takes the pattern matching
notation---\texttt{*} and \texttt{?} and bracket expressions---and
applies it to find matching files in the filesystem \iffull relative to the
shell's current working directory \fi (Figure~\ref{expeg:globbing}).
Double quotes prevent pathname expansion (as in the last line).

\emph{Quote removal} merits a single sentence in the POSIX
standard: ``The quote characters (<backslash>, single-quote, and
double-quote) that were present in the original word shall be removed
unless they have themselves been quoted''.
You can see this effect in the final line of Figure~\ref{expeg:globbing}.
\iffull
That is, single quotes are used to prevent \emph{all} expansion---all
characters in single quotes are treated as escaped. Double quotes
prevent field splitting and pathname expansion, i.e., they ensure that
the quoted string is treated as a single field.
There is one critical exception: \texttt{\textdollar{}\at}, when
quoted, will expand to exactly as many fields as there are positional
parameters.
\fi
\fi

\subsection{Smoosh: a foundational, formal interpretation of the POSIX specification}

Our work aims to offer (a) a formal interpretation of the
specification that is also (b) a conforming implementation.
Such a compromise is imperfect: we don't model all possible behaviors
in the specification, but rather \emph{interpret} the specification
deterministically. Our formal model is in code and not a tiny core
calculus.
Nevertheless, we believe our model is interesting and our
implementation is useful.
Analogously, the CompCert compiler doesn't conform to MSVC, ICC, GCC,
or Clang exactly, and yet its model of the C language can be seen as
canonical~\cite{Blazy2009,10.1007/978-3-319-08970-6_36}.
Building such a foundational model is a necessary first step towards
more minimal calculi and more useful tools in a setting where
implementations (currently and unyieldingly) disagree on a number of issues (see
Section~\ref{sec:beyond}).
\TODO{more discussion of formal vs. code?}

Our paper has three parts:
a formal description, to show that while the shell is complex, it is
amenable to conventional techniques (Sections~\ref{sec:syntax}
and~\ref{sec:semantics});
a description of our implementation, to give more concrete detail on
our semantics (Sections~\ref{sec:impl} and~\ref{sec:stepper});
and a discussion of conformance, showing that
we have adequately modeled the POSIX shell
(Section~\ref{sec:results}) and extensions (Section~\ref{sec:beyond}).

\iffull
We begin by describing the shell's semantics using the conventional
tools: an AST (Section~\ref{sec:syntax}, Figure~\ref{fig:syntax})
along with a notion of state (Section~\ref{sec:state},
Figure~\ref{fig:state}) and extra runtime terms
(Figure~\ref{fig:rtsyntax}) for a small-step operational semantics
(Section~\ref{sec:semantics}; rule excerpts for expansion in
Section~\ref{sec:expansion}'s Figures~\ref{fig:expansion},
\ref{fig:word-expansion}; rule excerpts for evaluation in
Section~\ref{sec:evaluation}'s Figures~\ref{fig:eval-negation},
\ref{fig:eval-simple-expand}, and~\ref{fig:eval-simple}.).
\fi

\definecolor{shellpink}{RGB}{242,128,133}

\newcommand{\tallphantom}{\vphantom{gG}}
\newcommand{\RT}[1]{\highlight[shellpink]{{#1}}}
\newcommand{\RTv}[1]{\highlight[shellpink]{\tallphantom{#1}}}

\section{Syntax}
\label{sec:syntax}

The shell has an idiosyncratic concrete syntax, which we mirror
in our abstract syntax (Figure~\ref{fig:syntax}) but do not follow
strictly.
Since we're giving abstract syntax, we ignore the details resolved during parsing, particularly backslash escapes of reserved symbols and
single quoting.
A few conventions: we use \texttt{fixed-width} fonts for system concepts (e.g.,
\fd) and for abstract syntax that mirrors the shell's concrete syntax
(e.g., \texttt{>|}); we use $\mathsf{sans\text{-}serif}$ fonts for Smoosh concepts and $\mathit{italics}$ for metavariables. We occasionally use
subscript ``hints'' to translate English names into concrete shell
syntax (e.g., $\mathsf{prefix}_\text{\#}$).
We write $x^?$ for optional nonterminals\iffull: $x^? ::= x \BNFALT
\bullet$ for all non-terminals $x$\fi. We use Kleene star $x^*$ to
represent lists\iffull: $x^* ::= \cdot \BNFALT x^* ~ x$ for all non-terminals
$x$; equivalently, $x^* ::= \vec{x}$. We use\else{} and \fi $x^+$ to 
represent non-empty lists.
We use $n$ to refer to natural numbers and $b$ to refer to
booleans, tagging such variables with a descriptive name. For example,
$n_\exitstatus$ is a number representing an exit status and $\bsubst$
is a boolean that indicates whether a command substitution has been
performed (which helps determine the exit status; see \rn{CmdAssignDoneNoCmd} in Figure~\ref{fig:eval-simple}).
\iffull
Finally, we occasionally use $\mid$ to signal alternation outside
of a BNF definition (e.g., $\mathit{js}$'s $\mathsf{stopped}$ entry in
Figure~\ref{fig:state}).
\fi

\begin{figure*}
\[ \begin{array}{lrcl}
  \text{Commands} & c & ::= & 
  ( s \texttt{=} w )^* ~ w ~ r^* \BNFALT
  \mathsf{pipe}_{\texttt{|}} ~ c^+ ~ \amp^? \BNFALT
  c ~ r^+ \BNFALT
  c ~ \amp \BNFALT
  \texttt{(} ~ c ~ \texttt{)} \BNFALT 
  c_1 \texttt{;} c_2 \BNFALT
  \\ 
  \multicolumn{4}{r}{
  c_1 \texttt{\&\&} c_2 \BNFALT
  c_1 \texttt{||} c_2 \BNFALT
  \texttt{!} c \BNFALT
  \texttt{while} ~ c_1 ~ c_2 \BNFALT
  \texttt{for} ~ s ~ w ~ c \BNFALT 
  \texttt{if} ~ c_1 ~ c_2 ~ c_3 \BNFALT
  \texttt{case} ~ w ~ \mathit{cb}^* \BNFALT
  s\texttt{()} ~ c
  }  \\

  \text{Redirections} & r & ::= & 
    \mathsf{file} ~ \fd ~ \mathit{ft} ~ w \BNFALT
    \mathsf{dup} ~ \fd ~ \mathit{dt} ~ w \BNFALT
    \mathsf{here} ~ \fd ~ \mathit{ht} ~ w
    \\
  \text{File redirections} & \mathit{ft} & ::= & 
    \texttt{>} \BNFALT 
    \texttt{>|} \BNFALT
    \texttt{<} \BNFALT
    \texttt{<>} \BNFALT
    \texttt{>>}
    \\
  \text{Dup redirections} & \mathit{dt} & ::= & \texttt{>\&} \BNFALT \texttt{<\&} \\
  \text{Heredoc redirections} & \mathit{ht} & ::= & \mathsf{default} \BNFALT \mathsf{noexpand} \\  

  \text{Case branches} & \mathit{cb} & ::= &  \texttt{(} \textit{w}^+ \texttt{)} ~ c \\

\fullbreak

  \text{Word} & w & ::= & ( s \BNFALT \FS \BNFALT k )^* \\
  \text{Control codes} & k & ::= & 
    \tilde{}s^? \BNFALT
    \param{s}{\phi} \BNFALT
    \backtick{c} \BNFALT
    \arith{w} \BNFALT
    \quote{w} \BNFALT \\

\fullbreak

  \text{Parameter formats} & \phi & ::= &
    \mathsf{normal} \BNFALT
    \mathsf{default}_{\mathit{null}\texttt{-}} ~ w \BNFALT
    \mathsf{assign}_{\mathit{null}\texttt{=}} ~ w \BNFALT
    \mathsf{error}_{\mathit{null}\texttt{?}} ~ w \BNFALT \\ &&&
    \mathsf{alt}_{\mathit{null}\texttt{+}} ~ w \BNFALT
    \mathsf{length}_{\texttt{\#}} \BNFALT
    \mathsf{sub} ~ \mathit{side} ~ \mathit{mode} ~ w \\
  \text{Treatment of null} & \mathit{null} & ::= & \mathsf{string} \BNFALT \mathsf{unset}_\texttt{:} \\
  \text{Substring side} & \mathit{side} & ::= & \mathsf{prefix}_{\texttt{\#}} \BNFALT \mathsf{suffix}_{\texttt{\%}} \\
  \text{Substring mode} & \mathit{mode} & ::= & \mathsf{shortest}_{\texttt{s}} \BNFALT \mathsf{longest}_{\texttt{ss}} \\

\fullbreak

  \text{Non-empty strings} & s & \in & \Sigma^+ \text{ (e.g., UTF-8)} \\
  \text{File descriptors} & \fd & \in & \mathbb{N} \\
\end{array} \]

\caption{The shell's source syntax}
\label{fig:syntax}
\end{figure*}

The two primary AST constructs are the \emph{command}, $c$, and the \emph{words}, $w$.
Commands are the top-level construct: a user enters a command $c$ which has words $w$; the words are \emph{expanded} and the command is eventually \emph{evaluated}.

The base case for the command AST is the \emph{simple command}.
Simple commands $( s \texttt{=} w )^* ~ w ~ r^*$ model
command invocations (i.e., builtins, functions, and executables), plain assignments, and plain redirections.
All other command forms are composite.
First, there are some ``modifiers'' of simple commands:
pipelines, $\mathsf{pipe}_{\texttt{|}} ~ c^+ ~ \amp^?$, which may be in the background;
redirected commands, $c ~ r^+$;
background commands a/k/a asynchronous commands, $c ~ \amp$;
and subshells, $\texttt{(} ~ c ~ \texttt{)}$, noting that curly braces are
used for disambiguating parsing.
There are sequencing commands and logic:
sequence, $c_1 \texttt{;} c_2$;
short-circuiting conjunction, $c_1 \texttt{\&\&} c_2$;
short-circuiting disjunction, $c_1 \texttt{||} c_2 $;
and negation, $\texttt{!} c$.
There are two iteration constructs---while loops, $\texttt{while} ~
c_1 ~ c_2$; and for loops, $\texttt{for} ~ s ~ w ~ c$---and two
conditional constructs---the conventional conditional, $\texttt{if} ~
c_1 ~ c_2 ~ c_3$; and string pattern matching, $\texttt{case} ~ w ~
\mathit{cb}^*$, where each \emph{case branch} $\mathit{cb}$ pairs one
or more patterns (as words $w$) with a command $c$ to run when a pattern matches.
Function definition, $s\texttt{()} ~ c$, is also a command.

Redirections $r$ come in three fundamental forms: to a file
($\mathsf{file}$), to an existing file descriptor ($\mathsf{dup}$),
and from a given string a/k/a heredoc ($\mathsf{here}$).
File redirections, $\mathsf{file} ~ \fd ~ \mathit{ft} ~ w$, specify
a source file descriptor, a file mode $\mathit{ft}$ (e.g., \texttt{$>$} to write to a file), and a file target $w$.
File descriptor redirections, $\mathsf{dup} ~ \fd ~ \mathit{dt}
~ w$, can copy and close file descriptors.\iffull\footnote{It may be
  surprising that, for conventional filesystems, the direction
  $\mathit{dt}$ doesn't matter: the $\mathsf{dup}$ redirections turn into
  \texttt{dup2} system calls, which have no notion of direction.}\fi{}
Finally, heredoc redirections, $\mathsf{here} ~ \fd ~ \mathit{ht} ~ w$,
expand words $w$ and make the resulting string available for reading
on a given file descriptor $\fd$.
Redirections are typically scoped---that is, they take effect only for
a given simple command or composite redirection command; the
\texttt{exec} special builtin suppresses scoping, causing the
redirection to take permanent, global effect in the shell (see
discussion of \helper{runCmd} in Section~\ref{sec:evaluation}).

A \emph{word} $w$ is the primary sub-part of commands. Words are subject to expansion, 
showing up in several places:
in the assignments and arguments of simple commands;
as the targets of redirections;
as the strings iterated over in \texttt{for} loops;
and as both the scrutinees and patterns of \texttt{case}
conditionals.
A word $w$ is a possibly empty list of one of:
a \emph{user string}---a non-empty string $s$ from some character
set, e.g., UTF-8, which we designate $\Sigma$;
a statically-parsed field separator $\FS$;
or, a \emph{control code}, $k$.
In our abstract syntax, escaped control codes---like the literal
\texttt{\textbackslash\textdollar}---are ordinary strings.
In the initial stages of word expansion, only control codes are
expanded.
You might expect \texttt{*} to appear as a control code, but pathname
expansion is a dynamic search on the results of expansion and is not statically parsed.

There are five control codes:
tilde-and-prefix, $\tilde{}s^?$, is used to refer to users' home
directories (noting that the tilde prefix $s^?$ may be empty);
parameters $\param{s}{\phi}$ take a variable name $s$ and a
\emph{parameter format} $\phi$ which will determine how the results of
looking up $s$ will be used;
command substitutions $\backtick{c}$ hold commands that ought to be
run with their output captured;
arithmetic substitutions $\arith{w}$ hold words $w$ that will be
further expanded and then parsed as an integer arithmetic expression;
and quotation $\quote{w}$ inhibits field splitting and
pathname expansion.

Parameter formats come in two flavors: they allow for convenient
defaulting when variables are unset or null (i.e., hold the empty
string); and they enable post-processing of the result of lookup.
The $\mathsf{normal}$ format is the default and performs
standard variable lookup.
There are four defaulting parameter formats: $\mathsf{default}$, $\mathsf{assign}$, $\mathsf{error}$, and
$\mathsf{alt}$.
All four take words $w$ and a flag $\mathit{null}$: when
$\mathit{null} = \mathsf{string}$, then a variable set to the empty
string is still considered set; when $\mathit{null} =
\mathsf{unset}_\texttt{:}$ (where \texttt{:} is a concrete syntax
hint), then a variable set to the empty string is considered unset.
When $\mathsf{default}$ is used on a variable considered unset
(because it's actually unset or because $\mathit{null} =
\mathsf{unset}_\texttt{:}$), then the words $w$ are returned for
further expansion (rather than the empty string).
When $\mathsf{assign}$ is used on a variable considered unset, the
words $w$ are returned for further expansion---and the result of
expanding those words is assigned to $s$.
When $\mathsf{error}$ is used on a variable considered unset, the
words $w$ are returned for further expansion---and then used as an
error message.
The $\mathsf{alt}$ format is the \emph{opposite} of
$\mathsf{default}$: when it is used on a variable considered unset,
the null string is returned; when it is used on a variable considered
set, the words $w$ are expanded and returned.
The two post-processing parameter formats are $\mathsf{length}_\texttt{\#}$ and $\mathsf{sub} ~ \mathit{side} ~ \mathit{mode} ~ w$.
The former calculates the length of the value of the given parameter; the
latter does $\mathsf{prefix}_{\texttt{\#}}$ or
$\mathsf{suffix}_{\texttt{\%}}$ removal using either a
$\mathsf{shortest}_{\texttt{s}}$ or $\mathsf{longest}_{\texttt{ss}}$
match policy on a given pattern $w$.

\iffull Several \else Beyond simplifying the concrete syntax and escaping, several \fi shell features are omitted from our abstract syntax: our
parser desugars \texttt{until} loops into \texttt{while} loops; the
tab-stripping heredoc redirection \texttt{<<-} is handled in the
parser.
\iffull
Other bits of concrete syntax are ignored: infix and postfix keywords
for conditionals and loops (e.g., \texttt{then}, \texttt{esac}); our
AST doesn't allow you to omit the first parenthesis in a case branch;
quoted heredocs (\texttt{<<"EOF"}) are marked
$\mathsf{noexpand}$, as they will not undergo expansion.
\fi

\section{Semantics}
\label{sec:semantics}

POSIX specifies a broad set of behaviors for the shell; our semantics
is not small. We show excerpts here of the shell's semantics to (a) show
that the shell is nevertheless amenable to standard techniques, and (b) give a
sense of the level of detail of our semantics.
This section is a selective transcription of Smoosh's
implementation (Section~\ref{sec:impl}), meant to communicate the core
ideas in a concise mathematical form without minutae (like logging or
symbolic value manipulation).
%
%
We \emph{do} show all of our state definitions
(Section~\ref{sec:state}, Figure~\ref{fig:state}) and all of the
intermediate forms used by our small step semantics
(Figure~\ref{fig:rtsyntax}).
We take care lay out these forms of state plainly, even if we do not give our entire semantics in
mathematical notation.

We explain expansion (Section~\ref{sec:expansion}) before evaluation
(Section~\ref{sec:evaluation}). Rather than give an up-front
description of every helper function and system call, we introduce
them as needed.

\subsection{State}
\label{sec:state}

\begin{figure*}
\[ \begin{array}{lrcl}
  \text{Shell state} & \sigma &::=& \langle 
    \pid_{\mathsf{root}},
    b_{\mathsf{outermost}},
    \mathsf{opts},
    \mathsf{jobs},
    \mathsf{traps},
    \mathsf{traps}^?_{\mathsf{supershell}}, \\ &&& \phantom{\langle}
    \rho,
    s^*_{\mathsf{\texttt{\textdollar{}*}}},
    \ell^*, 
    \mathcal{V}_{\mathsf{ro}},
    \mathcal{V}_{\mathsf{export}},
    \rho_f,
    \mathsf{aliases}, 
    \mathsf{locale}, \\ &&& \phantom{\langle}
    s_{\mathsf{cwd}},
    \pid_\mathsf{\texttt{\textdollar{}!}},
    n_{\exitstatus},
    n_{\mathsf{loop}},
    n^?_{\mathsf{optoff}}
    \rangle \\
    
\fullbreak
  \text{Shell options} & \mathsf{opts} &\in& 
    \mathsf{Opts} = \{ \texttt{allexport}, \dots \} \\
  
\fullbreak
  \text{Traps} & \mathsf{traps} &:& \mathsf{Sig} \rightharpoonup s \\

\fullbreak  

  \text{Jobs} & \mathsf{jobs} &:& 
    \mathit{id}\mathord{:}\mathbb{N} \rightharpoonup 
    \{ \mathsf{ji} \mid \mathsf{ji}.n_{\mathsf{id}} = \mathit{id} \} \\
  \text{Job info} & \mathsf{ji} &::=& 
    \langle n_{\mathsf{id}}, 
            (\pid ~ c)^*_{\mathsf{pipe}},
            \pid,
            c,
            \mathit{js} \rangle \\
  \text{Job status} & \mathit{js} &::=& 
    \mathsf{running} \BNFALT 
    \mathsf{stopped} ~ (\texttt{TSTP} \BNFALT 
    \texttt{STOP} \BNFALT \texttt{TTIN} \BNFALT \texttt{TTOU}) \BNFALT \\ &&&
    \mathsf{terminated} ~ \sig \BNFALT \mathsf{done} ~ n_{\exitstatus} \\
  \text{Signals} & \sig & \in & \mathsf{Sig} = \{ \mathsf{SIGHUP}, \dots \} \\

\fullbreak

  \text{Global environments} & \rho &:& s \rightharpoonup s \\
  \text{Local environments} & \ell &:& 
    s \rightharpoonup s^? \times b_{\mathsf{ro}} \times b_{\mathsf{export}} \\
  \text{Sets of variable names} & \mathcal{V} &\subseteq& \mathcal{P}({\Sigma^*}) \\
  \text{Function definitions} & \rho_f &:&
    s \rightharpoonup c \\
  \text{Aliases} & \mathsf{aliases} &:&
    s \rightharpoonup s \\
  \text{Locales} & \mathsf{locale} &\in& \mathcal{L} \text{ (e.g., \texttt{C}, \texttt{it\_IT.UTF-8})} \\
\end{array}
\]

  \caption{The shell's state}
  \label{fig:state}
\end{figure*}

The POSIX shell has to track a significant amount of state (Figure~\ref{fig:state}).
Each line of the description of Smoosh's shell state characterizes different levels of detail.
The first line is about high-level process info: what is the root process ID ($\pid_{\mathsf{root}}$, to be used for \texttt{\textdollar\textdollar}, even in subshells)? Is this the outermost shell or a subshell ($b_{\mathsf{outermost}}$)? What shell options are set ($\mathsf{opts}$)? What jobs are running ($\mathsf{jobs}$)?
Smoosh tracks not only the current shell's signal handlers a/k/a traps ($\mathsf{traps}$), but also any traps from a supershell ($\mathsf{traps}^?_{\mathsf{supershell}}$) in order to properly implement the \texttt{trap} special builtin.\iffull\footnote{The core issue is that one would like to be able to find out which traps are currently set without writing to a file, but \backtick{\texttt{traps --}} will execute in a subshell.}\fi

The second line of the shell's state characterizes the environment: $\rho$ is the global, dynamically scoped environment; the current positional parameters are stored in a list $s^*_{\mathsf{\texttt{\textdollar{}*}}}$.
In addition to environment and positional parameters, Smoosh also tracks a stack of local environments, $\ell^*$. These local environments exist not only to support the non-POSIX builtin \texttt{local}, but also for scoped assignments for function calls (see Section~\ref{sec:local}).
The shell tracks the read-only and exported variables in $\rho$ via $\mathcal{V}_{\mathsf{ro}}$ and $\mathcal{V}_{\mathsf{export}}$, respectively.
Aliases and the current locale information are also tracked. (Smoosh only supports the ambient locale, though; see Section~\ref{sec:limitations}.)

The third and final line of the shell's state holds on to finer grained
information: the current working directory ($s_{\mathsf{cwd}}$), the
PID of the last background command
($\pid_{\mathsf{\texttt{\textdollar{}!}}}$), the exit status of the
last command ($n_\exitstatus$), how deeply nested we are in the current loop ($n_{\mathsf{loop}}$), and the offset into the argument for parsing in \texttt{getopts} ($n^?_{\mathsf{optoff}}$) (see Section~\ref{sec:getopts}).

Smoosh's state is just one way to keep track of everything a shell needs to know.
For example, the entire last line of the shell state could instead be
kept in the shell's environment, $\sigma.\rho$. Some shell state in
fact \emph{must} be kept there, e.g., the \texttt{OPTIND} variable
used by the \texttt{getopts} builtin.
We find it more convenient to manually track the last exit status in
$\sigma.n_\exitstatus$ as a number than to pretend it is an actual
environment variable in $\sigma.\rho$.
The POSIX standard specifies the minimum state for a shell in \S{}2.12. Confusingly,
some of the specified state is kept by the OS, not the shell (e.g., ``open files
inherited upon invocation of the shell, plus open files controlled by
exec'').

Taking Smoosh's general approach as a given, there are still alternative designs. 
For example, we track an explicit stack of local variables in the
shell state $\sigma$, but we don't have such a stack for function
parameters: instead we use $\mathsf{call}$ stack frames to save the
old positional parameters (Figure~\ref{fig:rtsyntax}).
Not having a stack of positional parameters makes it easy to ensure
that we never have an underflow of the positional parameter stack;
having an explicit stack of locals makes it easier to do scoped lookup.
Our variable lookup routine works as follows:
it immediately returns the value for a special variable (e.g., the value of \texttt{\textdollar{}?} is stored in $\sigma.n_\exitstatus$) or positional parameter (in $\sigma.s^*_{\mathsf{\texttt{\textdollar{}*}}}$);
failing that, it traverses the stack of locals $\sigma.\ell^*$;
failing that, it checks the global environment $\sigma.\rho$.

{\iffull
The POSIX shell offers no real data structures to its users: variables hold strings.
Worse still, this state is dynamically scoped.
By default, the POSIX shell offers two forms of lexically scoped state: file descriptors (which are unique to each process and therefore to each subshell) and positional parameters (which are lexically scoped to functions).
In a formal sense, these forms of local state are `enough', and not
merely in the way that a two counter machine is `enough'. Modernish
uses file descriptors, FIFO pipes, and subshells to implement loops
that respect local variables; it's certainly less efficient than an in
process loop, but it allows a programmer to write reliably scoped code
on nearly any shell.
It may be a hair shirt, but it is at least warm one.
\fi}

\begin{figure*}
\[ \begin{array}{lrcl}
  \text{Commands} & c & ::= & \multicolumn{1}{r}{\dots \BNFALT 
    \mathsf{cmd}_{\mathsf{args}} ~ ( s \texttt{=} w )^* ~ \mathit{es} ~ r^* ~ co \BNFALT
    \mathsf{cmd}_{\mathsf{redirs}} ~ ( s \texttt{=} w )^* ~ f ~ \mathit{rs} ~ co \BNFALT} \\ 
  \multicolumn{4}{r}{
    \mathsf{cmd}_{\mathsf{assigns}} ~ ( s \texttt{=} \mathit{es} )^* ~ f ~ \mathit{sfds} ~ co 
    \mathsf{cmd}_{\mathsf{ready}} ~ \rho ~ s_{\mathsf{cmd}} ~ f_{\mathsf{args}} ~ \mathit{sfds} ~ co \BNFALT
    \mathsf{run} ~ \rho_{\mathsf{cmd}} ~ s_{\mathsf{cmd}} ~ f_{\mathsf{args}} ~ \mathit{sfds} ~ \mathit{co} \BNFALT} \\
  \multicolumn{4}{r}{
    \mathsf{while}_{\mathsf{cond}} ~ c_{\mathrm{orig}} ~ c_{\mathrm{cur}} ~ c_{\mathrm{body}} \BNFALT
    \mathsf{while}_{\mathsf{body}} ~ c_{\mathrm{cond}} ~ c_{\mathrm{body}} ~ c_{\mathrm{cur}} \BNFALT} \\
  \multicolumn{4}{r}{
    \mathsf{for}_{\mathsf{args}} ~ s ~ \mathit{es} ~ c \BNFALT
    \mathsf{for}_{\mathsf{start}} ~ s ~ f ~ c \BNFALT
    \mathsf{for}_{\mathsf{running}} ~ s ~ f ~ c_{\mathrm{body}} ~ c_{\mathrm{cur}} \BNFALT} \\
  \multicolumn{4}{r}{
    \mathsf{case}_{\mathsf{arg}} ~ \mathit{es} ~ \mathit{cb}^* \BNFALT
    \mathsf{case}_{\mathsf{match}} ~ s ~ \mathit{cb}^* \BNFALT
    \mathsf{case}_{\mathsf{check}} ~ s ~ \mathit{es}_{\mathsf{pat}} ~ c ~ \mathit{cb}^* \BNFALT} \\
  \multicolumn{4}{r}{
    \mathsf{call} ~ n_{\mathsf{loop}} ~ s^*_{\mathsf{\texttt{\textdollar{}*}}} ~ s_\mathsf{fun} ~ c_{\mathsf{orig}} ~ c_{\mathsf{cur}} \BNFALT
    \mathsf{break} ~ n \BNFALT \mathsf{continue} ~ n \BNFALT \mathsf{return} \BNFALT \mathsf{exit} \BNFALT \mathsf{done} \BNFALT
    \mathsf{redirs} ~ c ~ \mathit{sfds} \BNFALT } \\ 
  \multicolumn{4}{r}{
    \mathsf{eval} ~ n_{\mathsf{linno}} ~ \texttt{pctx} ~ s_\mathsf{src} ~ b_{\mathsf{interactive}} ~ b_{\mathsf{toplevel}} \BNFALT
    \mathsf{eval}_\mathsf{cmd} ~ n_{\mathsf{linno}} ~ \texttt{pctx} ~ s_\mathsf{src} ~ b_{\mathsf{interactive}} ~ b_{\mathsf{toplevel}} ~ c \BNFALT} \\ 
  \multicolumn{4}{r}{
    \mathsf{exec} ~ s_{\mathsf{path}} ~ s_{\mathsf{name}} ~ f_{\mathsf{args}} ~ \rho ~ b_{\mathsf{/bin/sh}} \BNFALT
    \mathsf{wait} ~ \pid ~ b_{\mathsf{checked}} ~ b_{\mathsf{cmd}} \BNFALT
    \mathsf{trapped} ~ \sig ~ n_{\exitstatus} ~ c_{\mathsf{handler}} ~ c_{\mathsf{cont}}} \\
  \text{Command options} & \mathit{co} & ::= & \langle \bsubst, b_{\mathsf{fork}}, b_{\mathsf{simple}} \rangle \\
  \text{Redirection state} & \mathit{rs} & ::= & \langle \mathit{er}^* ~ \mathit{xr}^? ~ \mathit{r}^* \rangle \\
  \text{Expanded redir} & \mathit{er} & ::= & 
    \mathsf{efile} ~ \mathit{ft} ~ \fd ~ s \BNFALT
    \mathsf{edup} ~ \mathit{dt} ~ \fd ~ \fd^? \BNFALT
    \mathsf{ehere} ~ \mathit{ht} ~ \fd ~ s
    \\
  \text{Expanding redir} & \mathit{xr} & ::= & 
    \mathsf{xfile} ~ \mathit{ft} ~ \fd ~ \mathit{es} \BNFALT
    \mathsf{xdup} ~ \mathit{ft} ~ \fd ~ \mathit{es} \BNFALT
    \mathsf{xhere} ~ \mathit{ht} ~ \fd ~ \mathit{es}
    \\

 \fullbreak

  \text{Control codes} & k & ::= & \dots \BNFALT 
    \mathsf{assign} ~ s ~ w \BNFALT
    \mathsf{error} ~ s ~ w \BNFALT 
    \mathsf{match} ~ f ~ \mathit{side} ~ \mathit{mode} ~ w \BNFALT \\
    &&&
    \mathsf{cmdsubst} ~ c ~ \pid ~ \fd \BNFALT
    \mathsf{cmdwait} ~ c ~ \pid ~ s
  \\

\fullbreak

  \text{Expansion state} & \mathit{es} & ::= &
    \mathsf{start} ~ \mathit{eo} ~ w \BNFALT
    \mathsf{expand} ~ \mathit{eo} ~ e ~ w \BNFALT
    \mathsf{split} ~ \mathit{eo} ~ e \BNFALT \\ &&& 
    \mathsf{path} ~ \mathit{eo} ~ i \BNFALT
    \mathsf{quote} ~ \mathit{eo} ~ i \BNFALT
    \mathsf{error} ~ f \BNFALT
    \mathsf{done} ~ f \\
  \text{Expansion options} & \mathit{eo} &::=&
    \langle b_{\mathsf{split}}, b_{\mathsf{glob}} \rangle \\

  \text{Expanded words} & e & ::= &
  ( \FS \BNFALT 
     \mathsf{src} ~ s \BNFALT
     \mathsf{exp} ~ s \BNFALT
     \texttt{@} ~ f \BNFALT
     \quote{s}
  )^* \\
  
  \text{Intermediate fields} & i & ::= &
  ( \mathsf{ws} ~ \FS \BNFALT 
     \FS \BNFALT 
     s \BNFALT
     \quote{s}
  )^* \\

  \text{Fields} & f & ::= & s^* \\
   
\fullbreak

  \text{Parsing context} & \texttt{pctx} & \in & \mathcal{L} \text{ (parser state)} \\
  \text{Saved FDs} & \mathit{sfds} & : & \fd \rightharpoonup \mathit{ofd} \\
  \text{Old FD info} & \mathit{ofd} & ::= & \mathsf{restore} ~ \fd \BNFALT \mathsf{close} \\
\end{array} \]

  \caption{Smoosh's runtime extensions for use in the small-step semantics}
  \label{fig:rtsyntax}
\end{figure*}

\subsection{Word expansion}
\label{sec:expansion}

\begin{figure*}
\definecolor{wcolor}{RGB}{192,192,192}
\definecolor{ecolor}{RGB}{249,115,250}
\definecolor{icolor}{RGB}{255,0,0}
\definecolor{fcolor}{RGB}{255,0,0}

\tikzstyle{stage}=[draw,rectangle,rounded corners,align=center,text=black]
\tikzstyle{words}=[stage,draw=none,fill=wcolor]
\tikzstyle{expwords}=[stage,ecolor,densely dotted,text=black]
\tikzstyle{ifields}=[stage,icolor,densely dashed,text=black]
\tikzstyle{fields}=[stage,fcolor,text=black]

\tikzstyle{trans}=[->,thick,rounded corners]

\begin{tikzpicture}[thick, double, node distance=3cm]
  \node[words] (start) 
       {$\mathsf{start} ~ \mathit{eo} ~ w$};
  \node[words] (expand) [right of=start,node distance=2.75cm] 
       {$\mathsf{expand} ~ \mathit{eo} ~ e ~ w$};
  \node[ifields] (pathname) [right of=expand,node distance=4cm]  
       {$\mathsf{pathname} ~ \mathit{eo} ~ i$};
  \node[expwords] (split) [below of=pathname, node distance=1cm] 
       {$\mathsf{split} ~ \mathit{eo} ~ e$};
  \node[ifields] (quote) [right of=pathname, node distance=4.25cm] 
       {$\mathsf{quote} ~ \mathit{eo} ~ i$};
  \node[fields] (error) [below of=expand, node distance=1cm]    
       {$\mathsf{error} ~ f$};
  \node[fields] (done) [below of=quote, node distance=1cm] 
       {$\mathsf{done} ~ f$};

  \draw[trans] (start) -- (expand);
  \draw[trans] (expand.east) -- ([xshift=0.5cm] expand.east) |- (split.west);
  \draw[trans,densely dashed] (expand) -- (pathname) 
    node[midway,above] {$\neg\mathit{eo}.b_{\mathsf{split}}$};
  \draw[trans] (expand) -- (error);
  \draw[trans] (split) -- (pathname);
  \draw[trans] ([yshift=0.15cm] pathname.east) -- ([yshift=0.15cm] quote.west);
  \draw[trans,densely dashed] ([yshift=-0.15cm] pathname.east) -- ([yshift=-0.15cm] quote.west)
    node[midway,below] {$\texttt{noglob} \vee \neg \mathit{eo}.b_{\mathsf{glob}}$};
  \draw[trans] (quote) -- (done);
\end{tikzpicture}

~ \\

\hdr{Word expansion}{\hfill
\fbox{$\sigma, \mathit{es} \stepsto \sigma, \bsubst, \mathit{es}$} 
}
  
{\infrule[ExpStart]
  {}
  {\sigma, 
   \mathsf{start} ~ \mathit{eo} ~ w
   \stepsto
   \sigma,
   \bot_\subst,
   \mathsf{expand} ~ \mathit{eo} ~ \cdot ~ w
  }
}

{\infrule[ExpExpand]
  {\sigma, 
   \langle \mathit{eo}.b_{\mathsf{split}}, \bot_{\mathsf{"}}, \bot_{\mathsf{gen}} \rangle,
   e,
   w
   \stepsto
   \sigma, \bsubst, \langle e', w' \rangle
  }
  {\sigma,
   \mathsf{expand} ~ \mathit{eo} ~ e ~ w
   \stepsto
   \sigma', \bsubst,
   \mathsf{expand} ~ \mathit{eo} ~ e' ~ w'
  }
}

{\infrule[ExpExpandErr]
  {\sigma, 
   \langle \mathit{eo}.b_{\mathsf{split}}, \bot_{\mathsf{"}}, \bot_{\mathsf{gen}} \rangle,
   e,
   w
   \stepsto
   \sigma, \bsubst, 
   \mathsf{error} ~ e_{\mathsf{err}}
  }
  {\sigma,
   \mathsf{expand} ~ \mathit{eo} ~ e ~ w
   \stepsto
   \sigma', \bsubst,
   \mathsf{error} ~ \mathsf{toFields}(e_{\mathsf{err}})
  }
}

{\infrule[ExpExpandSplit]
  {\mathit{eo}.b_{\mathsf{split}}}
  {\sigma,
   \mathsf{expand} ~ \mathit{eo} ~ e ~ \cdot
   \stepsto
   \sigma, \bot_\subst,
   \mathsf{split} ~ \mathit{eo} ~ e
  }
}

{\infrule[ExpExpandNoSplit]
  {\neg\mathit{eo}.b_{\mathsf{split}}}
  {\sigma,
   \mathsf{expand} ~ \mathit{eo} ~ e ~ \cdot
   \stepsto
   \sigma, \bot_\subst,
   \mathsf{path} ~ \mathit{eo} ~ \mathsf{skipSplitting}(e)
  }
}

{\infrule[ExpSplit]
  {}
  {\sigma,
   \mathsf{split} ~ \mathit{eo} ~ e
   \stepsto
   \sigma, \bot_\subst,
   \mathsf{path} ~ \mathit{eo} ~ \mathsf{fieldSplitting}(\sigma, e)}
}

{\infrule[ExpPath]
  {\texttt{noglob} \not\in \sigma.\mathsf{opts} \andalso \mathit{eo}.b_{\mathsf{glob}}}
  {\sigma,
   \mathsf{path} ~ \mathit{eo} ~ i
   \stepsto
   \sigma, \bot_\subst,
   \mathsf{quote} ~ \mathit{eo} ~ \mathsf{pathnameExpansion}(\sigma, i)   
  }
}

{\infrule[ExpPathNoGlob]
  {\texttt{noglob} \in \sigma.\mathsf{opts} \vee \neg \mathit{eo}.b_{\mathsf{glob}}}
  {\sigma,
   \mathsf{path} ~ \mathit{eo} ~ i
   \stepsto
   \sigma, \bot_\subst,
   \mathsf{quote} ~ \mathit{eo} ~ \mathsf{unescape}(i)   
  }
}

{\infrule[ExpQuote]
  {}
  {\sigma,
    \mathsf{quote} ~ \mathit{eo} ~ i
    \stepsto_\bot
    \sigma, \bot_\subst,
    \mathsf{done} ~ \mathsf{combineFields}(\mathsf{removeQuotes}(i))
  }
}

\[ \begin{array}{@{}r@{~}c@{~}ll@{}}
  \multicolumn{3}{@{}l}{\textbf{Helper function}} & \textbf{Description} \\ \hline
  \helper{skipSplitting}(e) &=& i               & \text{Convert expanded words to intermediate fields} \\
  \helper{toFields}(e) &=& f                   & \text{Convert expanded words to fields} \\
  \helper{fieldSplitting}(\sigma, e) &=& i     & \text{Break expanded words into intermediate fields } \\
  \helper{unescape}(i) &=& i                   & \text{Remove glob escape characters} \\
  \helper{pathnameExpansion}(\sigma, i) &=& i  & \text{Expand globs (e.g., \texttt{*} and \texttt{?})} \\
  \helper{removeQuotes}(i) &=& i               & \text{Remove quote characters/AST nodes} \\
  \helper{combineFields}(i) &=& f              & \text{Conjoin adjacent fields and whitespace} \\ 
\end{array} \]

  \caption{Small-step semantics for word expansion}
  \label{fig:expansion}
\end{figure*}

Word expansion occurs when evaluating assignments, commands, redirections, the
iteratee of a \texttt{for} loop, and both the scrutinee and the patterns of
\texttt{case} conditionals.
The general outline of the process takes a \emph{word} $w$ and expands it
to a list of strings, called \emph{fields} $f$.
We sketched the features of word expansion by example already (Section~\ref{sec:exp-informal});
we now give a formal model of word expansion.

We model expansion as a transition system between expansion states
$\mathit{es}$ (Figure~\ref{fig:rtsyntax}). There are seven such expansion states
(Figure~\ref{fig:expansion}), not to be confused with POSIX's
seven components of its four stages.
The states are:
the initial state, $\mathsf{start} ~ \mathit{eo} ~ w$;
initial word expansion, $\mathsf{expand} ~ \mathit{eo} ~ e ~ w$, where
we perform the first stage of POSIX word expansion (tildes,
parameters, command substitution, and arithmetic);
field splitting,
$\mathsf{split} ~ \mathit{eo} ~ e$;
pathname expansion,
$\mathsf{path} ~ \mathit{eo} ~ i$;
quote removal,
$\mathsf{quote} ~ \mathit{eo} ~ i$;
and two terminal states, one for errors, $\mathsf{error} ~ f$, and one
for successful completion, $\mathsf{done} ~ f$.
A diagram of the possible transitions
(Figure~\ref{fig:expansion}, top) shows that errors can
only occur during control code expansion; that field splitting can be
skipped wholesale (dashed line); and that pathname expansion is always
run, but sometimes it doesn't actually glob but rather unescapes
globbing characters (dashed line). The colors follow the legend in
Figure~\ref{fig:shtepper}.
We give a lower level understanding of the possible transitions
(Figure~\ref{fig:expansion}, bottom) with inference rules for a
small-step semantics, where $\sigma,
\mathit{es} \rightarrow \sigma', \bsubst, \mathit{es}$ means that in
shell state $\sigma$, the expansion state $\mathit{es}$ transitions to
a new shell state $\sigma'$ and a new expansion state $\mathit{es}$.
The boolean flag $\bsubst$ indicates whether or not a command
substitution was run.
The transition system conditions on the
expansion options $\mathit{eo}$.

\begin{figure*}
\hdr{Initial word expansion}{\hfill
\fbox{$\sigma, \mathit{wo}, e, w \stepsto \sigma, \bsubst, \mathit{wer}$} 
}

\flushleft
\[ \begin{array}{lrcl}
  \text{Initial word expansion options} & \mathit{wo} &::=&
    \langle b_{\mathsf{split}}, b_{"}, b_{\mathsf{gen}} \rangle \\
  \text{Initial word expansion result} & \mathit{wer} &::=&
    \langle e, w \rangle \BNFALT
    \mathsf{error} ~ e \\
\end{array} \]
\centering

\sidebyside[.45][.49]
{\infrule[EWSep]
  {}
  {\sigma, \mathit{wo}, e, \FS ~ w 
   \stepsto
   \sigma, \bot_\subst, \langle e ~ \FS, w \rangle
  }
}
{\infrule[EWLit]
  {\mathit{lit} = \begin{cases}
      \quote{s} & \phantom{\neg}\mathit{wo}.b_{\mathsf{"}} \\
      \mathsf{exp} ~ s & \neg \mathit{wo}.b_{\mathsf{"}} \wedge \phantom{\neg}\mathit{wo}.b_{\mathsf{gen}} \\
      \mathsf{src} ~ s &\neg \mathit{wo}.b_{\mathsf{"}} \wedge \neg\mathit{wo}.b_{\mathsf{gen}} \\
    \end{cases}
  }
  {\sigma, \mathit{wo}, e, s ~ w 
   \stepsto
   \sigma, \bot_\subst, \langle e ~ \mathit{lit}, w \rangle
  }
}

%
%

~ \\[1em]

\sidebyside[.48][.48]
{\infrule[EWCtrl]
  {\sigma, \mathit{wo}, k \stepsto \sigma', \bsubst, \langle e', w' \rangle}
  {\sigma, \mathit{wo}, e, k ~ w 
   \stepsto
   \sigma, \bsubst, \langle e ~ e', w' ~ w \rangle}
}
{\infrule[EWCtrlErr]
  {\sigma, \mathit{wo}, k \stepsto \sigma', \bsubst, \mathsf{error} ~ e}
  {\sigma, \mathit{wo}, e, k ~ w 
   \stepsto
   \sigma, \bsubst, \mathsf{error} ~ e}
}

~\\[.5em]

\hdr{Control-code expansion}{\hfill
\fbox{$\sigma, \mathit{wo}, k \stepsto \sigma', \bsubst, \mathit{wer}$} 
}

{\infrule[CmdSubst]
  {(\sigma_1, \fd_r, \fd_w) = \mathsf{pipe}(\sigma_0) \\
   (\sigma_2, \pid) = \mathsf{forkShell}(\sigma, c ~ \fd_w\texttt{>\&}1 ~ \fd_r\texttt{>\&}\bullet) \andalso
   \sigma_3 = \texttt{close}(\sigma_2, \fd_w)
  }
  {\sigma_0, \mathit{wo}, \backtick{c}
   \stepsto
   \sigma_3, \top_\subst, \langle \cdot, \mathsf{cmdsubst} ~ c ~ \pid ~ \fd_r \rangle
  }
}

{\infrule[CmdSubstRead]
  {(\sigma_1, s) = \texttt{readAll}(\sigma_0, \fd_r) \andalso \sigma_2 = \texttt{close}(\sigma_1, \fd_r)}
  {\sigma_0, \mathit{wo}, 
   \mathsf{cmdsubst} ~ c ~ \pid ~ \fd_r 
   \stepsto
   \sigma_2, \top_\subst, 
   \langle \cdot, \mathsf{cmdwait} ~ c ~ \pid ~ \mathsf{trimRNL}(s) \rangle
  }
}

{\infrule[CmdSubstReadErr]
  {(\sigma_1, \mathsf{error} ~ e) = \texttt{readAll}(\sigma_0, \fd_r)}
  {\sigma_0, \mathit{wo}, 
   \mathsf{cmdsubst} ~ c ~ \pid ~ \fd_r 
   \stepsto 
   \sigma_1, \top_\subst, \mathsf{error} ~ e
  }
}

{\infrule[CmdSubstWait]
  {(\sigma_1, n'_\exitstatus) = \texttt{wait}(\sigma_0, \pid)}
  {\sigma_0, \mathit{wo}, 
   \mathsf{cmdwait} ~ c ~ \pid ~ s
   \stepsto 
   \sigma_1[n_\exitstatus = n'_\exitstatus], \top_\subst, 
   \langle \mathsf{exp} ~ s, \cdot \rangle
  }
}
\[ \begin{array}{@{}r@{~}c@{~}ll@{}}
  \multicolumn{3}{@{}l}{\textbf{System call}} & \textbf{Description} \\ \hline
  \syscall{pipe}(\sigma) &=& (\sigma', \fd_r, \fd_w)                         & \text{Create a FIFO pipe} \\
  \syscall{forkShell}(\sigma, c) &=& (\sigma, \pid)                          & \text{Fork a new shell running $c$} \\
  \syscall{close}(\sigma, \fd) &=& \sigma                                    & \text{Close a file descriptor} \\
  \syscall{readAll}(\sigma, \fd) &=& (\sigma', s \BNFALT \mathsf{error} ~ e) & \text{Read all a file descriptor until EOF} \\
  \syscall{waitpid}(\sigma, \pid) &=& (\sigma', n_\exitstatus)                    & \text{Wait for a PID} \\

\\

  \multicolumn{3}{@{}l}{\textbf{Helper function}} & \textbf{Description} \\ \hline
  \helper{trimRNL}(s) &=& s' & \text{Remove trailing newlines} \\
\end{array} \]

  \caption{Small-step semantics for initial word expansion ($\mathsf{expand}$) and command substitution}
  \label{fig:word-expansion}
\end{figure*}

Initial word expansion in the expansion state $\mathsf{expand}$ is defined as a small-step operational
semantics $\sigma, \mathit{wo}, e, w \stepsto \sigma,
\bsubst, \mathit{wer}$, where the initial $e$ and $w$ represent the
current progress expanding the words $w$ into the \emph{expanded
  words} $e$ (or an error, as recorded in the \emph{intial word expansion result} $\mathit{wer}$).
We track whether or not a command substitution is performed
($\bsubst$), along with a variety of options particular to initial word
expansion ($\mathit{wo}$): whether splitting should happen (for a special case of \texttt{\textdollar{}*}), whether
the current control codes are quoted, and whether the current control
codes are from user input or are indirectly generated.

The rules map over each component of the words under expansion,
pulling an element off the front, processing it, and adding the result
to the already expanded words $e$.
We offer here only an excerpt of the control code expansion rules:
those dealing with command substitution
(Figure~\ref{fig:word-expansion}). Command substitution is exemplary
because (a) it ties the recursive knot between evaluation and
expansion, (b) it makes several system calls, and (c) its evaluation
produces more control codes.

How does command substitution work?
Given a control code $\backtick{c}$, we must run the command $c$,
capturing its output and saving it for further expansion.
The \rn{CmdSubst} rule starts the process.
First, we  use the \syscall{pipe} system call to create a POSIX
pipe, a kernel-managed FIFO queue with a file descriptor at each end:
one for reading ($\fd_r$) and one for writing ($\fd_w$).
Next, we fork a subshell, running our command $c$ with some
redirections: first, standard output is redirected to the write end of
our pipe, $\fd_w$; second, inside of the forked process, we close the
read end of our pipe, $\fd_r$.\iffull\footnote{If our outer shell were to die
  for some reason, then the nested process will be killed by a
  \texttt{SIGPIPE} when it writes to $\fd_w$.}\fi{}
Finally, we close the write end of the pipe in our running shell,
stepping to an intermediate $\mathsf{cmdsubst}$ form that tracks both the
$\pid$ of the forked subshell and the pipe's read
end.
With the command running, we try to read all of the file descriptor
(using the system call \texttt{readAll}; \rn{CmdSubstRead}). Once we've read everything,
we close $\fd_r$, save the string, and use the $\mathsf{cmdwait}$ form
to wait on the command to finish.
If reading produces an error, the entire expansion process errors
(\rn{CmdSubstReadErr}).
When the command has finished, we save its exit status and return
the command's output as the expanded words $\mathsf{exp} ~ s$
(\rn{CmdSubstWait}).
Note that each of the \rn{CmdSubst*} rules sets $\bsubst$ to
true\iffull---setting it once is all that is strictly necessary, but
we set it each time for clarity\fi.

\iffull
We could have written different rules, where we use the transitive
closure of evaluation to define all of command substitution in a
single rule.
Such a rule would be well defined: transitive closure is monotonic, so
we the least fixed point of our expansion/evaluation rules would still
exist. But such a rule would be inconvenient in our implementation
(where we want to carefully observe each of the steps) and in any use
of our metatheory for proof\iffull (where the use of transitive closure
seriously complicates inductive proofs)\fi.
\fi

\subsection{Evaluation}
\label{sec:evaluation}

Shell programs are evaluated in a line-oriented fashion:
a command is read and then immediately evaluated.
\iffull
What is a command in the shell? 
We informally group them into a few categories: execution, control, and function
definitions. Every command leaves behind an integer \emph{exit
  status} indicating whether the command succeeded (0) or failed
(non-zero).
Execution commands could be an assignment \texttt{x=5}, a simple command
\texttt{ls}, a backgrounded command \texttt{rm -r \_build/ \amp}, a redirected
command \texttt{find . -name \textbackslash{}*.c >src\_files}, a pipeline \texttt{cat *.txt | wc -l}, or a subshell \texttt{(echo \textdollar{}PPID)}.
Control commands could be sequence \texttt{cd src; make}, standard
boolean control (using an exit status of 0 to mean true) as in
\texttt{make || echo failed} or \texttt{! diff file1 file2} or
\texttt{kill -0 \textdollar{}P \&\& echo process is alive}, a boolean
conditional \texttt{if [ "\textdollar{}\#" -ne 1 ]; then echo expected
  one argument; exit 2; fi}, a pattern matching conditional
\texttt{case \textdollar{}(whoami) in (root) echo be careful!;; (*) echo user
  mode;; esac}, a conventional boolean loop \texttt{while [
    "\textdollar{}x" -lt 5 ]; do x=\textdollar((x+1)); done}, or an
expanding/iterating loop \texttt{for x in a b c; do echo
  \textdollar{}x; done}.
Finally, function definition is also a command \texttt{cleanup() \{ rm -r \textdollar{}TMP; \}}.
\TODO{explain each of these examples}

\fi
The most interesting bits of evaluation are simple commands $( s \texttt{=} w )^* ~ w_{\mathsf{cmd}} ~ r^*$, which make a
variety of system calls.
We give a
detailed, formal semantics for them in Section~\ref{sec:eval-commands},
but two other command forms merit discussion: the \texttt{case} conditional
and the \texttt{for} loop.

The \texttt{case} conditional has the form \texttt{case $w$ in
  ($\mathrm{pat}_1$) $s_1$;; $\dots$ esac}, where $w$ are words to
expand and then match against the patterns $\mathrm{pat}_i$. The
`branch-on-pattern' style of conditional is a cousin of the
\texttt{switch} conditional seen in imperative languages like C,
though the shell uses string pattern matching instead of equality.
Case analysis doesn't involve so many steps, but it involves a
restricted form of expansion (stopping at arithmetic expansion) and
pattern matching; the quoting rules for patterns have subtle
interactions with the way dynamic pattern escaping (as opposed to parse-time escaping) is treated in the shell.

The \texttt{for} loop has the form \texttt{for $x$ in $w$; do $c$; done}, where $x$ is a variable name (without the \texttt{\textdollar}), $w$ are words to be expanded, and $c$ is a command. The words $w$ are expanded to zero or more fields; $c$ is executed once for each such field, with $x$ bound to each field in turn.
The shell's \texttt{for} loop is like a \texttt{for-each} loop in Java
or a \texttt{for ... of} loop in JavaScript, where the iterated over words
are interpreted as a list according to field splitting and
\verb|$IFS|.
%

\begin{figure*}
\hdr{Command semantics (negation)}{\hfill \fbox{$\sigma, \bcheck, c \stepsto \sigma, c$}}

\sidebyside[.35][.62]
{\infrule[Not]
  {\sigma, \top_\checkmark, c \stepsto \sigma', c'}
  {\sigma, \bcheck, 
   \texttt{!} c
   \stepsto
   \sigma', 
   \mathsf{!} c'}}
{\infrule[NotCtrl]
  {c \in \{ \mathsf{break} ~ n, \mathsf{continue} ~n, \mathsf{return}, \mathsf{exit} \}}
  {\sigma, \bcheck, 
   \texttt{!} c
   \stepsto
   \sigma, 
   c}}

~\\[.5em]

\sidebyside[.52][.45]
{\infrule[NotSuccess]
  {\sigma.n_\exitstatus = 0}
  {\sigma, \bcheck, 
   \texttt{!} \mathsf{done}
   \stepsto
   \sigma[n_\exitstatus = 1], 
   \mathsf{done}}}
{\infrule[NotFail]
  {\sigma.n_\exitstatus \ne 0}
  {\sigma, \bcheck, 
   \texttt{!} \mathsf{done}
   \stepsto
   \sigma[n_\exitstatus = 0], 
   \mathsf{done}}}

\caption{Small-step semantics for evaluation of negation}
\label{fig:eval-negation}
\end{figure*}

We give a semantics to commands as a small-step relation $\sigma,
\bcheck, c \stepsto \sigma', c'$, which we read as ``in state
$\sigma$, when $\bcheck$ indicates whether someone else is responsible
for checking our exit status, the command $c$ steps to a new state
$\sigma'$ and a new command $c'$''.
Much of the evaluation relation is standard: control (\texttt{;},
\texttt{if}, \texttt{\&\&}, \texttt{||}, \texttt{!},
\texttt{while}, function definition) works more or less the usual way,
though rather than having actual boolean values, the last exit status is
looked up in the shell state.

As a warmup, we give the four rules for negation
(Figure~\ref{fig:eval-negation}).
The first rule is a standard congruence rule: the term $\texttt{!} c$
takes a step by stepping $c$ itself (\rn{Not}). Note that we set
$\top_\checkmark$ in the premise. When the shell has the flag
\texttt{errexit} set, we will exit on a non-zero exit status, but that
behavior is proscribed ``when executing the compound list following
the \texttt{while}, \texttt{until}, \texttt{if}, or \texttt{elif}
reserved word, a pipeline beginning with the \texttt{!} reserved word,
or any command of an AND-OR list other than the
last''~(\cite{POSIXbase}, \S{}4, \texttt{set}). Setting $\top_\checkmark$ tells subsidiary rules that someone else will be checking the exit status, and the shell should not exit on error.
The second rule propagates control: if the command $c$ is one of the
internal AST nodes used to represent control, then we must propagate
the control command through the negation (\rn{NotCtrl}).
The final two rules do the actual work of negation, turning a
successful exit status of 0 into 1 (\rn{NotSuccess}) and an
unsuccessful exit status other than 0 into a 0 (\rn{NotFail}).
We show these four rules to assuage concerned readers: much of the semantics is standard.

Before offering a complete formal semantics for simple commands
(Section~\ref{sec:eval-commands}), we comment on some of the more
interesting runtime forms necessary in Smoosh's semantics
(Figure~\ref{fig:rtsyntax}).
The shell has a variety of line-oriented read/eval loops: the outermost shell itself, the `dot' (\texttt{.}) a/k/a \texttt{source}
command, and the \texttt{eval} command. In order to nest such loops, read/eval loops are part of Smoosh's AST:
$\mathsf{eval}$ nodes representing an eval loop reading commands and
$\mathsf{eval_{cmd}}$ nodes representing an eval loop executing a command.
There are several other interesting AST forms: $\mathsf{redirs}$ holds
on to information for restoring redirections; $\mathsf{exec}$ turns into a
call to \texttt{execve}; $\mathsf{wait}$ turns into a call to
\texttt{waitpid}; and $\mathsf{trapped}$ and $\mathsf{call}$ are stack
frames representing the current trap being processed and the current
function being called, respectively.

\subsubsection{Simple Commands}
\label{sec:eval-commands}

Here we highlight one of the key parts of the shell's semantics: the evaluation of simple commands
(Figures~\ref{fig:eval-simple-expand}, \ref{fig:eval-simple-helpers},
and~\ref{fig:eval-simple}).
Assignments, commands, and redirections are agglomerated in the POSIX
specification into a single form: a \emph{simple command} of the form
$( s \texttt{=} w )^* ~ w_{\mathsf{cmd}} ~ r^*$, where the $s$ are variable names,
each $w$ is a word for expansion either in an assignment or as the
command and arguments itself ($w_{\mathsf{cmd}}$), and each $r$ is a redirection.
How does the specification explain them?
Simple commands expand arguments, redirections, and then assignments---though assignments may be expanded before redirections
if there are no arguments (i.e., $w_{\mathsf{cmd}} = \cdot$).
A five-step, twelve-point outline determines how to perform assignments;
a two-step, eleven-point outline determines how to find out if a command should be resolved as a special builtin, a builtin, a function, or an executable.
Simple commands hardly live up to their name: they are not simple---in
fact, they are the most complex part of command evaluation---and they
may not even run a command!
%
 
\begin{figure*}
\hdr{Command semantics (simple commands)}{\hfill 
\fbox{$\sigma, \bcheck, c \stepsto \sigma, c$}}

\infrule[CmdStart]
  {}
  {\sigma, \bcheck, ( s \texttt{=} w )^* ~ w ~ r^* 
   \stepsto
   \sigma, ( s \texttt{=} w )^* ~ 
           (\mathsf{start} ~ \langle \top_{\mathsf{split}}, \top_{\mathsf{glob}} \rangle ~ w) ~ 
           r^* ~ \langle \bot_{\mathsf{cmdsubst}}, \top_{\mathsf{fork}}, \bot_{\mathsf{simple}} \rangle
  }

\infrule[CmdArgExp]
  {\sigma, \mathit{es} \stepsto \sigma', \bsubst, \mathit{es}' \andalso 
    \mathit{es}' \ne \mathsf{error} ~ f \andalso \mathit{es}' \ne \mathsf{done} ~ f}
  {\sigma, \bcheck, 
   ( s \texttt{=} w )^* ~ \mathit{es} ~ r^* ~ \mathit{co}
   \stepsto
   \sigma', 
   \mathsf{cmd}_{\mathsf{args}} ~ ( s \texttt{=} w )^* ~ \mathit{es}' ~ r^* ~ \mathsf{cmdSubst}(\mathit{co}, \bsubst)}

  {\infrule[CmdArgErr]
     {\sigma, \mathit{es} \stepsto \sigma', \mathsf{error} ~ f}
     {\sigma, \bcheck, 
      \mathsf{cmd}_{\mathsf{args}} ~ ( s \texttt{=} w )^* ~ \mathit{es} ~ r^* ~ \mathit{co}
      \stepsto
      \expError(\sigma', f, \top_{\mathsf{exit}})}}
  {\infrule[CmdArgDone]
     {\sigma, \mathit{es} \stepsto \sigma', \mathsf{done} ~ f}
     {\sigma, \bcheck, 
      \mathsf{cmd}_{\mathsf{args}} ~ ( s \texttt{=} w )^* ~ \mathit{es} ~ r^* ~ \mathit{co}
      \stepsto
      \sigma', 
      \mathsf{cmd}_{\mathsf{redirs}} ~ ( s \texttt{=} w )^* ~ f ~ \langle \cdot, \bullet, r^* \rangle ~ \mathit{co}}}

~ \\

{\infrule[CmdRedir]
  {\sigma, \mathit{rs} \stepsto \sigma', \bsubst, \mathit{rs}'}
  {\sigma, \bcheck, 
   \mathsf{cmd}_{\mathsf{redirs}} ~ ( s \texttt{=} w )^* ~ f ~ \mathit{rs} ~ \mathit{co}
   \stepsto
   \sigma, 
   \mathsf{cmd}_{\mathsf{redirs}} ~ ( s \texttt{=} w )^* ~ f ~ \mathit{rs}' ~ \mathsf{cmdSubst}(\mathit{co}, \bsubst)}}
{\infrule[CmdRedirErr]
  {\sigma, \mathit{rs} \stepsto \sigma', \mathsf{error} ~ f}
  {\sigma, \bcheck, 
   \mathsf{cmd}_{\mathsf{redirs}} ~ ( s \texttt{=} w )^* ~ f ~ \mathit{rs} ~ \mathit{co}
   \stepsto
   \expError(\sigma', f, \mathsf{special}(f)_{\mathsf{exit}})}}

{\infrule[CmdRedirDoneErr]
  {\mathsf{redir}(\sigma, \mathit{er}^*) = (\sigma', f) \\
    \begin{array}{r@{~}c@{~}l}  b_{\mathsf{exit}} &=&
     (\neg \bcheck \wedge \texttt{errexit} \in \sigma'.\mathsf{opts}) \\ &\vee&
     (\mathsf{special}(f_{\mathsf{cmd}}) \wedge 
      \neg \mathit{co}.b_{\mathsf{simple}} \wedge
      \neg \mathsf{interactive}(\sigma')))
   \end{array}
  }
  {\sigma, \bcheck, 
   \mathsf{cmd}_{\mathsf{redirs}} ~ ( s \texttt{=} w )^* ~ f_{\mathsf{cmd}} ~ \langle \mathit{er}^*, \bullet, \cdot \rangle  ~ \mathit{co}
   \stepsto
   \mathsf{redirError}(\sigma', f, b_{\mathsf{exit}})}}

{\infrule[CmdRedirDone]
  {\mathsf{redir}(\sigma, \mathit{er}^*) = (\sigma', \mathit{sfds})}
  {\sigma, \bcheck, 
   \mathsf{cmd}_{\mathsf{redirs}} ~ ( s \texttt{=} w )^* ~ f ~ \langle \mathit{er}^*, \bullet, \cdot \rangle  ~ \mathit{co}
   \stepsto {} \\
   \sigma'[\ell^* = \sigma'.\ell ~ \cdot], 
   \mathsf{cmd}_{\mathsf{assigns}} ~ ( s \texttt{=} \mathsf{start} ~ w ~ \langle \bot_{\mathsf{split}}, \bot_{\mathsf{glob}} \rangle )^* ~ f ~ \mathit{sfds} ~ \mathit{co}}}

{\infrule[CmdAssign]
  {\sigma, \mathit{es} \stepsto \sigma', \bsubst \mathit{es}' \andalso 
    \mathit{es}' \ne \mathsf{error} ~ f \andalso \mathit{es}' \ne \mathsf{done} ~ f}
  {\sigma, \bcheck,
   \mathsf{cmd}_{\mathsf{assigns}} ~ s \texttt{=} \mathit{es} ~ ( s \texttt{=} \mathit{es}'' )^* ~ f_{\mathsf{cmd}} ~ \mathit{sfds}  ~ \mathit{co}
   \stepsto {} \\
   \sigma,
   \mathsf{cmd}_{\mathsf{assigns}} ~ s \texttt{=} \mathit{es} ~ ( s \texttt{=} \mathit{es}'')^* ~ f_{\mathsf{cmd}} ~ \mathit{sfds}  ~ \mathsf{cmdSubst}(\mathit{co}, \bsubst)
  }}
{\infrule[CmdAssignErr]
  {\sigma, \mathit{es} \stepsto \sigma', \bsubst, \mathsf{error} ~ f}
  {\sigma, \bcheck,
   \mathsf{cmd}_{\mathsf{assigns}} ~ s \texttt{=} \mathit{es} ~ ( s \texttt{=} \mathit{es}'' )^* ~ f_{\mathsf{cmd}} ~ \mathit{sfds}  ~ \mathit{co}
   \stepsto
   \expError(\sigma, f, \top_{\mathsf{exit}})}}
{\infrule[CmdAssignSetErr]
  {\sigma, \mathit{es} \stepsto \sigma', \bsubst, \mathsf{done} ~ f \andalso
   \setLocal(\sigma', s, \toString(f)) = \mathsf{error} ~ f_{\mathsf{err}} \\
   b_{\mathsf{exit}} = (\neg \bcheck \wedge \texttt{errexit} \in \sigma'.\mathsf{opts}) \vee \neg \mathsf{interactive}(\sigma')
  }
  {\sigma, \bcheck,
   \mathsf{cmd}_{\mathsf{assigns}} ~  s \texttt{=} \mathit{es} ~ ( s \texttt{=} \mathit{es}'' )^* ~ f_{\mathsf{cmd}} ~ \mathit{sfds}  ~ \mathit{co}
   \stepsto
   \expError(\sigma, f_{\mathsf{err}}, b_{\mathsf{exit}})}}
{\infrule[CmdAssignSet]
  {\sigma, \mathit{es} \stepsto \sigma', \bsubst, \mathsf{done} ~ f \andalso
   \setLocal(\sigma', s, \toString(f)) = \sigma''
  }
  {\sigma, \bcheck,
   \mathsf{cmd}_{\mathsf{assigns}} ~ s \texttt{=} \mathit{es} ~ ( s \texttt{=} \mathit{es}'' )^* ~ f_{\mathsf{cmd}} ~ \mathit{sfds}  ~ \mathit{co}
   \stepsto {} \\
   \sigma'',
   \mathsf{cmd}_{\mathsf{assigns}} ~ ( s \texttt{=} \mathit{es}'' )^* ~ f_{\mathsf{cmd}} ~ \mathit{sfds}  ~ \mathit{co}
  }}

  \caption{Small-step semantics for simple commands (part 1: expansion)}
  \label{fig:eval-simple-expand}
\end{figure*}

In our model, evaluation of the simple command $( s \texttt{=} w )^* ~ w_\mathsf{cmd} ~ r^*$ proceeds as follows.
First, the words of the command arguments ($w_\mathsf{cmd}$) are expanded from left to right
(\rn{CmdStart}, \rn{CmdArgExp}, \rn{CmdArgErr}, \rn{CmdArgDone}).
Next, redirections are expanded (\rn{CmdRedir}, \rn{CmdRedirDoneErr})
and applied (\rn{CmdRedirDone}), saving file descriptor information $\mathit{sfds}$
for later unwinding.
Then we expand the words in the assignments (\rn{CmdAssign},
\rn{CmdAssignErr}) and store the results in a fresh local environment
(\rn{CmdAssignSet}).
At thus point, we are already in dangerous territory: the POSIX
specification allows assignments to be expanded before redirections
when there are no arguments. Our deterministic semantics always does
redirections first; \bash and \ksh will run assignments first
when there is no command.

\begin{figure*}
\[ \begin{array}{@{}r@{~}c@{~}ll@{}}
  \multicolumn{3}{@{}l}{\textbf{Helper function}} & \textbf{Description} \\ \hline
  \helper{special}(f) &=& b                                                    & \text{Determine if $f$ is a special builtin} \\
  \setParam(\sigma, x, s) &=& \sigma' \BNFALT \mathsf{error} ~ s               & \text{Set a parameter} \\
  \setLocal(\sigma, x, s) &=& \sigma' \BNFALT \mathsf{error} ~ s               & \text{Set a local parameter in the outermost scope} \\
  \helper{cmdSubst}(\mathit{co}, \bsubst) &=& \mathit{co}'                     & \text{Disjunctively update  $\mathit{co}.\bsubst$} \\
  \expError(\sigma, f, b_{\mathsf{exit}}) &=& (\sigma', c)                       & \text{Possibly exit with an expansion error} \\
  \helper{redirError}(\sigma, f, b_{\mathsf{exit}}) &=& (\sigma', c)             & \text{Possibly exit with a redirection error} \\
  \helper{redir}(\sigma, \mathit{er}^*) &=& (\sigma', \mathit{sfds} \BNFALT f_\mathsf{err}) & \text{Perform a redirection, recording saved FDs} \\
  \helper{unredir}(\sigma, \mathit{sfds}) &=& \sigma'                          & \text{Restore saved FDs} \\
  \multicolumn{3}{@{}l}{\helper{runCmd}(\sigma, \bcheck, \rho_{\mathsf{cmd}}, s, f, \mathit{co})} 
  & \text{Resolve command (builtin, function, executable)} \\
  &=& \multicolumn{2}{l@{}}{(\sigma', c, b_{\mathsf{restore}}) \BNFALT (\sigma', \mathsf{error} ~ f_{\mathsf{err}})} \\
  \helper{toAssigns}(\ell) &=& \rho                                                       & \text{Convert a local environment to bindings} \\
  \helper{mayExit}(b_{\mathsf{exit}}, c) &=& c'                                             & \text{Conditionally exit} \\
  \checkTraps(\sigma, c) &=& (\sigma', c')                                                & \text{Check for pending signals and load trap handlers} \\

&&& \\
  \multicolumn{3}{@{}l}{\textbf{System calls for \helper{runCmd}}} & \textbf{Description} \\ \hline
  \syscall{fileExists}(\sigma, s) &=& b & \text{Determine if a file exists} \\
  \syscall{fileExecutable}(\sigma, s) &=& b & \text{Determine if a file is executable} \\
  \multicolumn{3}{@{}l}{\syscall{execve}(\sigma, s_\mathsf{cmd}, s_\mathsf{argv[0]}, s^*_\mathsf{argv}, \rho, b_\mathsf{sh})} & \text{Replace the current process} \\
  &=& (\sigma', c \BNFALT \mathsf{error} ~ s)  & \\

&&& \\
  \multicolumn{3}{@{}l}{\textbf{System calls for \helper{redir} and \helper{unredir}}} & \textbf{Description} \\ \hline
  \syscall{fileRedir}(\sigma, \mathit{ft}, s) 
    &=& (\sigma', \fd \BNFALT \mathsf{error} ~ s) 
    & \text{Open a file redirection} \\
  \syscall{closeAndSaveFD}(\sigma, \fd) 
    &=& (\sigma', \mathit{sfds} \BNFALT \mathsf{error} ~ s) 
    & \text{Close an \fd, saving it at an $\fd \ge 10$} \\
  \multicolumn{3}{@{}l}{\syscall{renumberFD}(\sigma, b_\mathsf{close}, \fd_\mathsf{orig}, \fd_\mathsf{wanted})} & \text{Renumber an \fd, saving the target} \\
    &=& (\sigma', \mathit{sfds} \BNFALT \mathsf{error} ~ s) 
    &  \\
  \syscall{heredoc}(\sigma, s) 
    &=& (\sigma', \fd \BNFALT \mathsf{error} ~ s)   
    & \text{Create an \fd holding heredoc contents} \\

&&& \\
  \multicolumn{3}{@{}l}{\textbf{System calls for \helper{checkTraps}}} & \textbf{Description} \\ \hline
  \syscall{pendingSignal}(\sigma) &=& (\sigma', \sig^?) & \text{Get a signal if one is pending} \\

\end{array}\]

  \caption{Helpers for simple command execution}
  \label{fig:eval-simple-helpers}
\end{figure*}

Even in defining the relatively simple behavior of these expansions,
there's been call for a variety of helper functions
(Figure~\ref{fig:eval-simple-helpers}).
Some of these helpers are quite simple---$\helper{cmdSubst}$ just
updates the command options $\mathit{co}$ to reflect whether expansion performed command
substitution.
Others have significant logic and make system calls:
$\helper{runCmd}$ is 65 SLOC of Lem code that includes executable
resolution using \verb|$PATH| (via a separate, 50 SLOC function), a shell fork,
and an $\mathsf{exec}$ form which will cause the semantics to call
\syscall{execve}. The redirection functions $\helper{redir}$ and
$\helper{unredir}$ make a variety of system calls for manipulating
files ($\syscall{fileRedir}$, $\syscall{closeAndSaveFD}$,
$\syscall{renumberFD}$, and $\syscall{heredoc}$).

\begin{figure*}[t]
\hdr{Command semantics (simple commands, continued)}{\hfill \fbox{$\sigma, \bcheck, c \stepsto \sigma, c$}}

{\infrule[CmdAssignDoneNoCmd]
  {\sigma.\ell^* = \ell^*_0 ~ \ell_{\mathsf{cmd}} \andalso
    n'_{\exitstatus} = \begin{cases}
      \sigma.n_{\exitstatus} & \mathit{co}.\bsubst \\
      0 & \text{otherwise} \\
    \end{cases} \\
   \sigma' = \mathsf{unredir}(\sigma, \mathit{sfds})[\ell^* = \ell^*_0][\rho = \sigma.\rho[\ell_{\mathsf{cmd}}][n_{\exitstatus} = n'_{\exitstatus}]
  }
  {\sigma, \bcheck,
   \mathsf{cmd}_{\mathsf{assigns}} ~ \cdot  ~ \cdot ~ \mathit{sfds}  ~ \mathit{co}
   \stepsto
   \mathsf{xtrace}(\sigma', \ell^*_0),
   \mathsf{done}
  }}

{\infrule[CmdAssignDoneCmd]
  {\sigma.\ell^* = \ell^*_0 ~ \ell_{\mathsf{cmd}}}
  {\sigma, \bcheck,
   \mathsf{cmd}_{\mathsf{assigns}} ~ \cdot  ~ s ~ f ~ \mathit{sfds}  ~ \mathit{co}
   \stepsto {} \\
   \mathsf{xtrace}(\sigma, \ell_{\mathsf{cmd}} ~ s ~ f),
   \mathsf{cmd_{\mathsf{ready}}} ~ \mathsf{toAssigns}(\ell_{\mathsf{cmd}}) ~ s ~ f ~ \mathit{sfds} ~ \mathit{co}
  }}

{\infrule[CmdRunNoexec]
  {\texttt{noexec} \in \sigma.\mathsf{opts}}
  {\sigma, \bcheck, 
   \mathsf{cmd_{\mathsf{ready}}} ~ \rho ~ s_{\mathsf{cmd}} ~ f ~ \mathit{sfds} ~ \mathit{co}
   \stepsto
   \sigma, 
   \mathsf{done}}}

{\infrule[CmdRunSpecial]
  {\texttt{noexec} \not\in \sigma.\mathsf{opts} \andalso
   \mathsf{special}(s) \andalso \neg \mathit{co}.b_{\mathsf{simple}} \\
   \sigma' = \sigma[\rho = \sigma.\rho \cup \rho_{\mathsf{cmd}}]
  }
  {\sigma, \bcheck, 
   \mathsf{cmd_{\mathsf{ready}}} ~ \rho_{\mathsf{cmd}} ~ s ~ f ~ \mathit{sfds} ~ \mathit{co}
   \stepsto
   \sigma', \mathsf{run} ~ \rho_{\mathsf{cmd}} ~ s ~ f ~ \mathit{sfds} ~ \mathit{co}
  }
}

{\infrule[CmdRun]
  {\texttt{noexec} \not\in \sigma.\mathsf{opts} \andalso
   \neg \mathsf{special}(s) \vee \mathit{co}.b_{\mathsf{simple}}
  }
  {\sigma, \bcheck, 
   \mathsf{cmd_{\mathsf{ready}}} ~ \rho_{\mathsf{cmd}} ~ s ~ f ~ \mathit{sfds} ~ \mathit{co}
   \stepsto
   \sigma, \mathsf{run} ~ \rho_{\mathsf{cmd}}  ~ s ~ f ~ \mathit{sfds} ~ \mathit{co}
  }
}

{\infrule[RunFail]
  {\helper{runCmd}(\sigma, \bcheck, \rho_{\mathsf{cmd}}, s, f, \mathit{co}) = (\sigma', \mathsf{error} ~ f_{\mathsf{err}}) \\
   b_{\mathsf{exit}} = 
    (\neg \bcheck \wedge \texttt{errexit} \in \sigma'.\mathsf{opts}) \vee
    (\mathsf{special}(s) \wedge \neg \mathit{co}.b_{\mathsf{simple}} \wedge
     \neg \mathsf{interactive}(\sigma'))}
  {\sigma, \bcheck, 
   \mathsf{run} ~ \rho_{\mathsf{cmd}} ~ s ~ f ~ \mathit{sfds} ~ \mathit{co}
   \stepsto
   \sigma',
   \checkTraps(\mathsf{mayExit}(b_{\mathsf{exit}}, \mathsf{redirs} ~ \mathsf{done} ~ \mathit{sfds}))
  }
}

{\infrule[RunErr]
  {\helper{runCmd}(\sigma, \bcheck, \rho_{\mathsf{cmd}}, s, f, \mathit{co}) = (\sigma', c, b_{\mathsf{restore}}) \\
    \sigma'.n_{\exitstatus} \ne 0 \andalso
    \neg \bcheck \andalso 
    \texttt{errexit} \in \sigma'.\mathsf{opts}
  }
  {\sigma, \bcheck, 
   \mathsf{run} ~ \rho_{\mathsf{cmd}} ~ s ~ f ~ \mathit{sfds} ~ \mathit{co}
   \stepsto
   \sigma', \mathsf{exit}}
}

{\infrule[Run]
  {\helper{runCmd}(\sigma, \bcheck, \rho_{\mathsf{cmd}}, s, f, \mathit{co}) = (\sigma', c, \top_{\mathsf{restore}}) \\
    \sigma'.n_{\exitstatus} = 0 \vee \neg (\neg \bcheck \wedge \texttt{errexit} \in \sigma'.\mathsf{opts}) \andalso
    c' = \begin{cases}
      \mathsf{redirs} ~ c ~ \mathit{sfds} & b_{\mathsf{restore}} \\
      c & \text{otherwise} \\
    \end{cases}
  }
  {\sigma, \bcheck, 
   \mathsf{run} ~ \rho_{\mathsf{cmd}} ~ s ~ f ~ \mathit{sfds} ~ \mathit{co}
   \stepsto
   \sigma', c'}
}

  \caption{Small-step semantics for simple commands (part 2: evaluation)}
  \label{fig:eval-simple}
\end{figure*}

Resuming our explanation of simple commands: after all of the expansion has been done (Figure~\ref{fig:eval-simple-expand}), it is time to try to actually run the command (Figure~\ref{fig:eval-simple}). 
It may be the case that our simple command has no command at all, in
which case we pop the local assignments off the stack
and add them to the global environment $\sigma'.\rho$
(\rn{CmdAssignDoneNoCmd}).
The redirections are undone via \helper{unredir}.
Now $\bsubst$ pays off: if a command substitution was performed, we
should return its exit status in $\sigma'.n_\exitstatus$; if not, we
should succeed with exit status zero.
If, on the other hand, there is a command, then we step to a `ready' state,
capturing the assignments accumulated in the local environment.
Our use of \setLocal is a particular choice, not at all mandated by
POSIX, which doesn't have local variables. We use local environments 
because (a) we need quasi-local things to track assignments, and (b)
it is convenient to reuse the architecture we built for local variables
(see Section~\ref{sec:local}).

When faced with $\mathsf{cmd}_{\mathsf{ready}}$, there are three possibilities:
we could have disabled execution (\rn{CmdRunNoexec});
we could be running a special builtin, in which case POSIX mandates
that our assignment be global (\rn{CmdRunSpecial});
or we could be running an ordinary command (\rn{CmdRun}).
In the latter two cases, we step to the $\mathsf{run}$ form to
actually run a command.

Finally, the $\mathsf{run}$ form triggers a call to the $\helper{runCmd}$
helper function. One of three things happens:
we encounter an error of some kind, in which case noninteractive shells may exit (\rn{RunFail});
the command runs and produces a non-zero exit status, in which case
the may exit due to the \texttt{errexit} option (\rn{RunErr});
or, the command may run with no need to exit, in which case we continue with the command returned from \helper{runCmd} (\rn{Run}).
The new command $c$ yielded by \helper{runCmd} will
be one of three things:
$\mathsf{done}$, because a builtin was run to completion;
a $\mathsf{call}$ form, because a function was called;
or a $\mathsf{wait}$ form, because an executable was forked but should
run in the foreground, and so it needs to be waited on.
These forms will in almost all cases be wrapped with a
$\mathsf{redirs}$ form; the \texttt{exec} special builtin is an
exception, since its redirections should have global effect.
The $\mathsf{redirs}$ form will call \helper{unredir} to restore the
saved file descriptors $\mathit{sfds}$, as was done in
\rn{CmdAssignDoneNoCmd}.

It is the shell's responsibility to (a) receive and record signals as they arrive so that (b) those pending signals can be periodically checked and handled. We check for traps after completing the evaluation of commands.
The \helper{checkTraps} helper function uses the
\syscall{pendingSignal} system call to track these signals and
interrupt the main line of execution with trap handlers when necessary.
The $\mathsf{traps}$ field of the shell state tracks a partial
function from signals to strings. Signals not in the domain of
$\sigma.\mathsf{traps}$ have default dispositions; signals that map to
empty strings are ignored; otherwise, the strings are parsed and
interpreted when signals are handled.
Of the three \rn{Run*} rules, only one rule actually checks traps: \rn{RunFail}.
\rn{RunErr} need not check traps because the shell is exiting.
\rn{Run} need not check traps because all three of the possible forms
from \helper{runCmd} will check traps when they complete.

\section{Smoosh's implementation}
\label{sec:impl}

Having sketched in mathematical notation the Smoosh semantics, we turn our attention to Smoosh's actual executable semantics, i.e., the code.
Smoosh is written primarily in Lem~\cite{Mulligan:2014:LRE:2628136.2628143}, an ML-like language that extracts to OCaml and Coq, among others.
Smoosh uses \dash's parser via
\texttt{libdash},\blindurl{https://github.com/mgree/libdash} a
library that tracks \dash's main
repository\iffull\footnote{\url{https://git.kernel.org/pub/scm/utils/dash/dash.git/}}\fi{}
but has hooks for accessing the parser and a few other functions,
along with OCaml bindings to the \dash AST.
Smoosh is implemented across 20 files comprising 10\,814 SLOC, of
which 9\,305 lines are in Lem and 1\,509 are in OCaml. By way of
comparison, \dash has 14\,633 SLOC, of which 13\,318 are C code, 122
are shell scripts, and 1\,193 are header files.
The Smoosh code contains considerable extra material (about 1k SLOC):
shim code for reading the \dash AST and for rendering Smoosh ASTs in
JSON for the stepper (Section~\ref{sec:stepper}).
Smoosh is open source under the MIT license.\blindurl{https://github.com/mgree/smoosh} 

\iffull
\begin{table}[t]
  \begin{tabular}{@{}r@{\quad}llr@{}}
  & \textit{File} & \textit{Description} & \textit{SLOC} \\ \cmidrule[2pt]{2-4}

  \textbf{Drivers} 
  & \texttt{shell.ml} & Top-level shell & 216 \\
  & \texttt{shtepper.ml} & Symbolic stepper & 88 \\ 
  \cline{4-4} &&& \fpeval{216+88} \\

  &&& \\

  \textbf{Semantics}
  & \texttt{semantics.lem} & Small-step expansion and evaluation & 1249 \\
  & \texttt{command.lem} & Builtins, command resolution & 2210 \\
  & \texttt{fields.lem} & Field splitting and manipulation & 188 \\
  & \texttt{path.lem} & Pathname expansion & 79 \\ 
  & \texttt{arith.lem} & Arithmetic expansion & 586 \\
  & \texttt{pattern.lem} & Pattern matching (\texttt{case}, parameter formats) & 310 \\
  & \texttt{test.lem} & \texttt{test} builtin & 258 \\
  \cline{4-4} &&& \fpeval{1249 + 2210 + 188 + 79 + 586 + 310 + 258} \\

  &&& \\
  
  \textbf{OS typeclass}
  & \texttt{os.lem} & OS interface, FS definitions & 1130 \\
  & \texttt{os\_symbolic.lem} & Symbolic OS instance & 630 \\
  & \texttt{os\_system.lem} & Real OS instance & 348 \\
  & \texttt{system.ml} & Real OS helpers & 497 \\
  \cline{4-4} &&& \fpeval{1130 + 630 + 348 + 497} \\

  &&& \\
  \textbf{Definitions}
  & \texttt{smoosh.lem} & Top-level module & 29 \\
  & \texttt{smoosh\_prelude.lem} & AST, helpers pretty printing & 1898 \\
  & \texttt{signal.lem} & Signal definitions & 296 \\
  & \texttt{signal\_platform.lem} & Platform-specific signal information (generated) & 63 \\
  & \texttt{smoosh\_num.lem} & Numeric operations & 251 \\
  & \texttt{shim.ml} & Parsing and JSON serialization & 708 \\
  & \texttt{version.lem} & Version info (generated) & 7 \\
  \cline{4-4} &&& \fpeval{29 + 1898 + 296 + 63 + 251 + 708 + 7} \\
 

  \end{tabular}
  \caption{Breakdown of Smoosh's components}
  \label{tab:sloc}
\end{table}
\fi

Smoosh's architecture consists of a core semantics with two configurable `ports': one for the \emph{driver}, which determines how the Smoosh semantics are used; and one for the \emph{OS implementation}, which determines what system calls do and how the filesystem behaves. 
The core semantics is a small-step operational semantics, as sketched
in Section~\ref{sec:semantics}. The semantics itself is not
outrageously large (\iffull{}Table~\ref{tab:sloc}\else{}4\,880 SLOC for the semantics and
builtins, 2\,605 SLOC for the OS interface,
and 3\,252 for AST
definitions\fi), with a roughly 3:1 ratio of supporting code to code in the two core
stepping functions:
\begin{ocaml}
val step_expansion : forall 'a. OS 'a => 
  os_state 'a * expansion_state -> expansion_step * os_state 'a * expansion_state 
val step_eval : forall 'a. OS 'a => 
  os_state 'a -> checking_mode -> stmt -> evaluation_step * os_state 'a * stmt
\end{ocaml}
The \code{step_expansion} function (328 SLOC, the core expansion semantics) corresponds to the relation $\sigma, \mathit{es}
\stepsto \sigma, \bsubst, \mathit{es}$ (Section~\ref{sec:expansion});
the \code{step_eval} function (706 SLOC, the core evaluation semantics) corresponds to the relation $\sigma,
\bcheck, c \stepsto \sigma, c$ (Section~\ref{sec:evaluation}).
In addition to semantics for the POSIX shell language itself, we also provide implementations for all of the special builtins,\footnote{
\texttt{break}, 
\texttt{:}, 
\texttt{continue}, 
\texttt{.} a/k/a \texttt{source}, 
\texttt{eval}, 
\texttt{exec}, 
\texttt{exit}, 
\texttt{export}, 
\texttt{readonly}, 
\texttt{return}, 
\texttt{set}, 
\texttt{shift}, 
\texttt{times}, 
\texttt{trap}, and 
\texttt{unset}~\cite{POSIXbase} \S{}2.14}
mandatory builtins,\footnote{
\texttt{alias},
\texttt{bg}, 
\texttt{cd}, 
\texttt{command}, 
\texttt{false}, 
\texttt{fc}, 
\texttt{fg}, 
\texttt{getopts}, 
\texttt{hash}, 
\texttt{jobs}, 
\texttt{kill}, 
\texttt{newgrp}, 
\texttt{pwd}, 
\texttt{read}, 
\texttt{true}, 
\texttt{umask}, 
\texttt{unalias}, and
\texttt{wait}~\cite{POSIXbase} \S{}2.9.1 part 1(d)}
and several others.\footnote{\texttt{echo}, \texttt{help}, \texttt{history},
  \texttt{local}, \texttt{printf}, \texttt{test} a/k/a \texttt{[},
    \texttt{type}.}

\graf{The OS typeclass}

The most interesting aspect of the Smoosh implementation is its
\texttt{OS} typeclass.
We have already introduced several system calls in the formal model (Figures~\ref{fig:word-expansion} and~\ref{fig:eval-simple-helpers}).
The system calls described there are part of 40 different calls that can be made to the operating system.
We can break the OS interface into three areas of interest:
true system calls (14 functions), used to work with processes, signals, and job control;
file system calls (24 functions), used to traverse and manipulate the
file system, where 10 of these calls correspond to POSIX \texttt{stat}
and \texttt{lstat};
and parser interactions (2 functions), used to communicate values of
\texttt{PS1} and \texttt{PS2} to the \texttt{libdash} parser.
Note that the calls described here don't necessarily correspond
to POSIX-defined functions or any particular operating system's
interface. Some system calls correspond clearly (e.g., \syscall{execve}); 
other Smoosh system calls (e.g., \syscall{heredoc})
correspond to several system calls (create a pipe, possibly
spawn a process if the heredoc string is bigger than the OS buffer
size, write the heredoc text to the pipe).
Some other system calls don't correspond to true ``system calls'' in
any sense at all: \syscall{pendingSignal} doesn't actually make any
system calls, but merely looks at a data structure. We put
\syscall{pendingSignal} in the OS typeclass because the nature of that
signal-tracking data structure depends on the OS typeclass instance.

We have implemented two instances of the OS typeclass:
the \texttt{system} implementation makes real system calls to the
host OS, allowing us to use Smoosh as a real shell; 
the \texttt{symbolic} implementation makes no real system calls to the host OS but instead simulates a POSIX OS and filesystem. We
use the \texttt{symbolic} instance in our stepper
(Section~\ref{sec:stepper}).
\iffull
We can imagine other OS instances, e.g., a \texttt{readonly}
instance that allows one to read the filesystem but not write or call
\texttt{execve}.
\fi
The OS typeclass forces us to confront difficult issues in program
structure and API design. We can see the root of the issue in
\syscall{waitpid}.
Different implementations of \syscall{waitpid} must do drastically
different things.
In \texttt{system} mode, we expect \syscall{waitpid} to actually
call \texttt{wait4} or some similar host OS function, blocking until
the given PID has terminated.
In \texttt{symbolic} mode, we expect the current symbolic process to
suspend while another symbolic process proceeds.
That is, in \texttt{system} mode the OS typeclass is just a shim
between Smoosh and the host OS, and we can rely on the host OS's
scheduler entirely.
In \texttt{symbolic} mode, the OS typeclass must actually implement the
  scheduler itself; see Section~\ref{sec:scheduling}, ``Scheduling''.

\iffull
\graf{Challenges with Lem}
We chose to implement Smoosh in Lem so we could write one
implementation that extracts to OCaml for execution and to Coq for
proof.
In retrospect, it would have perhaps been wiser to simply directly
implement Smoosh in Coq. Coq can extract to OCaml, too, and while Lem
is well implemented, Coq is more robust and better supported. For
example, we had to extend Lem to support character literals in pattern
matching. When two branches of a pattern match disagree in return
type, Lem simply gives the line numbers for the whole match---not
particularly helpful for, e.g. \texttt{step\_eval}, an 894-line match!

It's not possible to use a typeclass while instantiating it in Lem, so
the actual \syscall{waitpid} system call takes \code{step_eval} as an
argument. The \texttt{system} mode has no need for \code{step_eval},
but the \texttt{symbolic} mode relies on it (see
Section~\ref{sec:scheduling}).
Our solution to this problem lead to some redundancy: each
instantiation of a typeclass defines the critical functions outside
the instance, with whatever dependency or mutual recursion is
necessary between operations. The typeclass instance definition then
simply references these external definitions.
Using Coq would not have helped with this issue: Coq does not allow
you to use an instance while defining it, either.
\fi

\subsection{Limitations}
\label{sec:limitations}

We have not attempted to implement any of the POSIX locale functionality.
It is a longer term goal to give a precise formal account of locales.
For now, however, Smoosh uses OCaml functions for ordering strings and
formatting numbers, which are locale-independent as of 2018. 
Smoosh's handling of the various terminal/TTY functions is incomplete\iffull
(see Section~\ref{sec:discussion}, ``Interactivity'')\fi.

\graf{Parsing}

Smoosh currently supports only the \texttt{libdash} parser, as
extracted from \dash.
The dash parser is certainly ``good enough'', as \dash is the default
\texttt{/bin/sh} on Debian and Ubuntu systems. Using its parser in Smoosh has
some drawbacks, though.
First, prompting using \texttt{PS1} and \texttt{PS2} is built in to
\dash's parser. These prompt variables ought to be subject to variable
expansion each time they are displayed---and \dash will (incorrectly)
use its own, internal expansion routine and environment to expand
these variables.
Second, the \dash parser doesn't support common extensions to the
POSIX spec, like statically parsed tests with the \texttt{[[} form.
Third, \dash's lexer doesn't correctly support multi-byte characters\iffull,
making it difficult to work with character sets like Unicode.\else, because \fi \dash's parser doesn't use
structured terms to represent words but instead inserts literal
control codes with negative signed character values\iffull (e.g., $-127$, or
$129$ as an unsigned number in two's complement, is an escape of the
next character)\fi.
Fourth and finally, \dash's parser is written in C and may therefore result in memory errors or security vulnerabilities.
We are interested in connecting Smoosh to other parsers, like
Morbig~\cite{regisgianas:hal-01890044} and the bash-compatible parser
from \OSH~\cite{Oil}.

\section{Smoosh's shell stepper}
\label{sec:stepper}

We have used the Smoosh semantics to implement a \emph{program
  stepper}, which we call the Shtepper.
The Shtepper traces shell programs in a simulated environment.
We use lightweight symbolic execution: symbolic values are
treated symbolically as much as possible, but we don't support
branches, i.e., we only explore a single path.
The Shtepper is implemented in two parts: a tracer and a visualizer.
The tracer takes a shell program
and a description of the environment in which to run it and produces a trace in JSON.
The visualizer takes a JSON trace and displays it in a browser using
JavaScript.
The Shtepper is publicly available online.\blindurl{http://shell.cs.pomona.edu/shtepper}
The tracer comprises 1.2k SLOC of Lem and OCaml, implementing a
synthetic POSIX environment (processes and filesystem), JSON
mappings for the Smoosh AST, and a driver for the semantics.
The visualizer comprises 1.5k SLOC of Ruby, ERB templates, JavaScript,
and CSS; the bulk of the code (1.2k SLOC) is in the JavaScript trace
rendering logic.

\begin{figure}[t]
  \includegraphics[width=.74\linewidth]{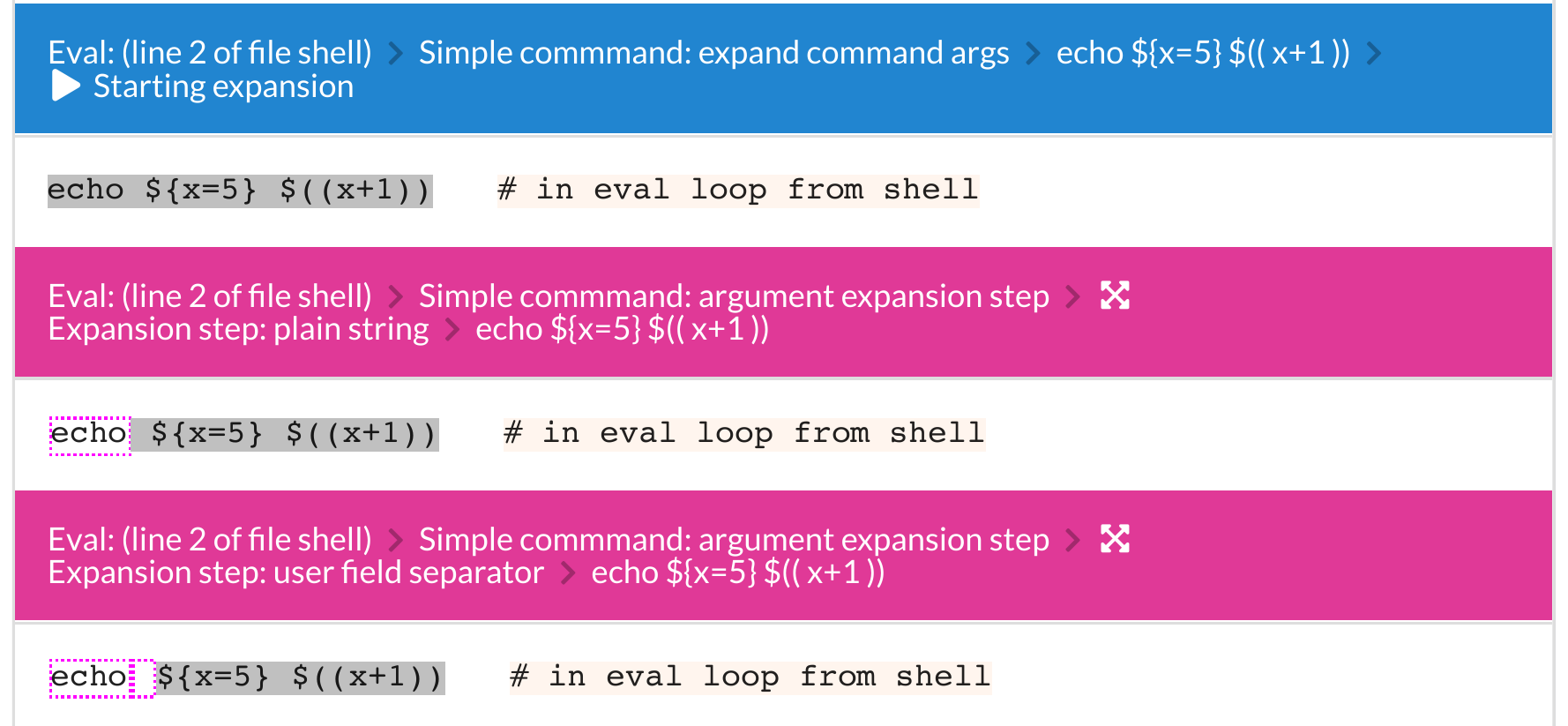}
  ~
  \includegraphics[width=.24\linewidth]{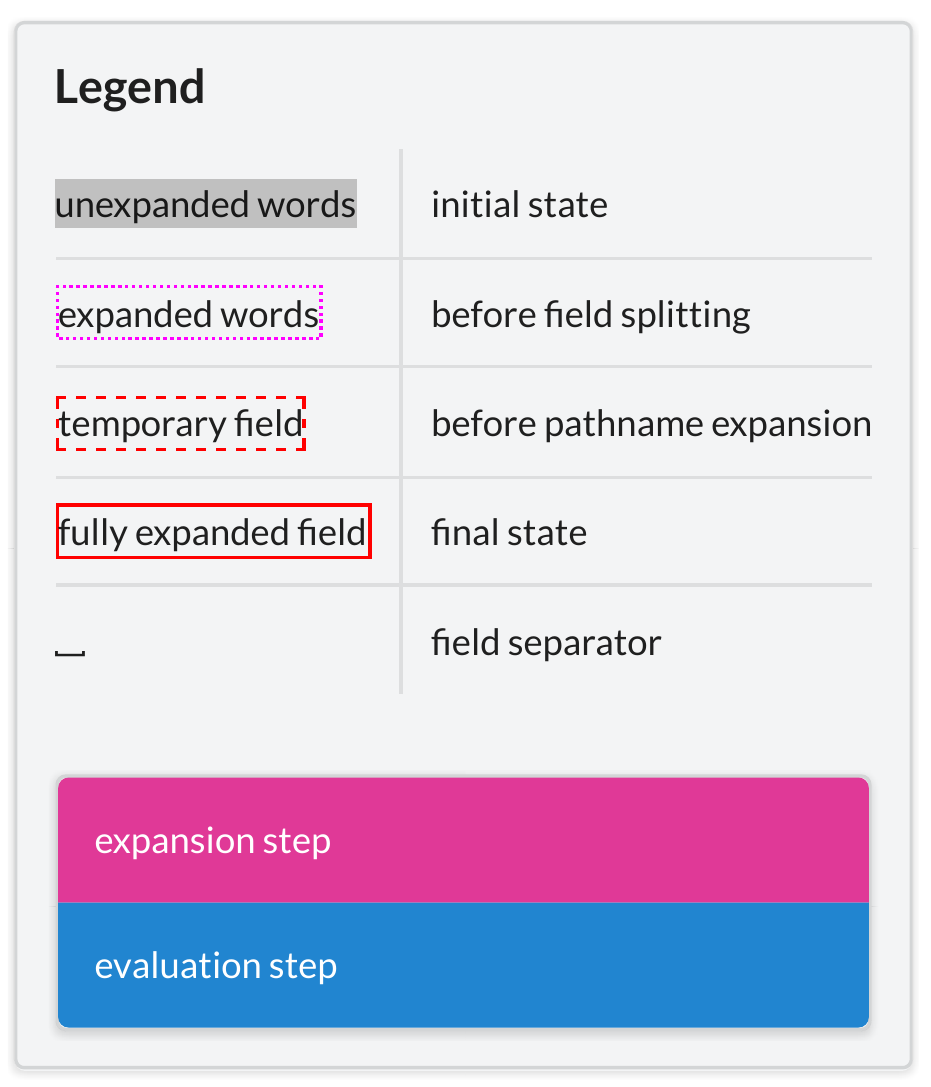}

  \caption{A fragment of a trace from the Shtepper}
  \label{fig:shtepper}
\end{figure}

\graf{The visualizer}

The Shtepper shows the main shell's thread of execution, highlighting
evaluation steps in {\color[RGB]{41,135,205} blue} and expansion
steps in {\color[RGB]{222,62,150} pink} (Figure~\ref{fig:shtepper}).
The shell contains a great deal of state and performs its work in many small steps.
It is difficult to know what to highlight, what to merely show, and what to hide.
By default, we show each step of expansion and evaluation, showing the
environment, STDOUT, and STDERR at the beginning, after changes,
and at the end of execution.
\iffull
The visualizer could of course be significantly improved: interactive
hide/show toggles for various bits of information, better ways to
specify the initial environment, initial environment presets, and so
on.
\fi
We plan to study which information best helps novices learn the shell and avoid pitfalls.

\subsection{Simulating POSIX}

Our stepper runs in a simulated POSIX environment, using the \texttt{symbolic} OS instance.
We use `fuel' to limit the extent of symbolic execution. The
web interface puts a conservative limit on fuel; at the risk of
nontermination, the fuel can be made infinite in the tracer.
\iffull
We are \emph{not} doing true symbolic execution, as we do not explore
every branch.
\fi

The symbolic shell state tracks the entirety of the POSIX environment:
a list of FIFO pipes;
a list of processes;
the root of the symbolic filesystem;
the current shell's mapping of file descriptors to pipes and files;
whether or not the current shell has exited;
the current shell's umask;
and, the contents of \texttt{/etc/passwd} to simulate calls to
\texttt{getpwnam} during tilde expansion.
Our symbolic model is sufficient to run interesting
pipes between subshells, but insufficient to run executables.
\iffull
At present, we implement two kinds of processes: subshells and zombies.
\fi
\TODO{IF TIME just hack in a simple thing for symbolic Exec}
\iffull
Subshells track their current status (running or stopped), the command they are executing, their internal shell state, and their file descriptors.
Zombies just hold on to their exit statuses.
\fi

\graf{Scheduling}
\label{sec:scheduling}

When simulating a symbolic POSIX system, which processes get to run when?
There is a tension between faithfully modeling all possible
interleavings of processes and offering concise, legible information
to the user of the stepper.
To motivate the question, consider the following two pipelines:
\texttt{while true; do echo 5; done | true} and \texttt{while true; do
  echo 5; done | \{ read x; echo \$((x+42)); \}}.
Both pipelines spawn two processes, both of which use shell builtins
exclusively: neither of these pipelines needs to make an
\texttt{execve} system call (though some systems may implement
\texttt{true} or \texttt{echo} as executables, Smoosh and most shells
build them in).
In both cases, the first process will send \texttt{5} on STDOUT
infinitely many times.
In the first pipeline, the second process ignores its input entirely,
terminating immediately\iffull with a successful exit status\fi.
In the second pipeline, the second process reads a line of
input, emits the line \texttt{47}, and terminates.

In a real POSIX system, both processes will be scheduled concurrently in both pipelines. The \texttt{while} loop will write as fast as it can until the pipe between the two processes is full, at which point the looping process will block.
The first pipeline's second process will terminate nearly immediately;
the second pipeline's second process will terminate right after it's
been able to read one line of input from the pipe.
In either case, when the second process terminates, the pipe between
the two processes has no more readers, and a \texttt{SIGPIPE} will be
sent to the first process, terminating it.

How should we simulate these pipelines?
We could use real threads to run the simulation, resulting in a
different schedule each time. Such an approach would be error prone
and nondeterministic; it is more appealing to use deterministic,
simulated threads.
How, then, do we schedule our simulated threads?
If we always run the pipeline left to right, then the pipelines above
will block when the pipe buffers fill up (or diverge if they're unbounded).
If we always run the pipeline right to left, then the first pipeline
above will terminate immediately but the second one will block,
waiting to read from a process that's never scheduled.
We need to do \emph{some} scheduling, but how?

We prioritize determinism and clarity over faithfully exploring all
possible schedules.
Scheduling takes place in rounds; every process in the system will
take a step every round (which may mean blocking).
Our scheduler will try to step each process in creation order: the
root shell will get to go first, and then the first subshell, and so
on.
But: whenever a shell waits on another process (e.g., because of a
foreground command or a command substitution) or would block reading
from a pipe, we take the opportunity to step the waited on or writing
process, making the scheduling order \emph{demand-driven}.
Scheduling according to demand keeps our scheduling deterministic
without rigidly following creation order.

When waiting for a command to terminate, we put no bound on the number
of steps we're willing to wait---blocking is the right
behavior.\footnote{Note that this waiting still counts as fuel usage,
  and our symbolic execution may give up while waiting.}
When a process calls \syscall{waitpid} in \texttt{symbolic} mode, the
waited-on process is looked up and stepped.
For a blocking wait---like \syscall{waitpid} or
\syscall{readAll}---our scheduling works well.
When waiting for a writer to supply input, we allow batches of 10
steps at a time, since one need not read all of the input: the
\texttt{read} builtin only needs to read a line, and we'll get a
shorter trace if we step the processes only enough to produce as much
output as we need.

\iffull 
Concretely, the \texttt{read} builtin uses the \syscall{readline}
system call:
\syscall{readLine}(\sigma, \fd, b_\mathsf{\texttt{\textbackslash}}) = 
(\sigma', s \BNFALT \mathsf{blocked} ~ \pid \BNFALT \mathsf{error} ~ s).$
A call to \syscall{readLine} tries to read from the file descriptor
\fd, with $b_\mathsf{\texttt{\textbackslash}}$ specifying whether a backslash
can allow a line continuation. The Smoosh \syscall{readLine} makes a
series of calls to the POSIX \texttt{read} primitive in
\texttt{system} mode.
Getting back a new state $\sigma'$, there are three possible
results: getting a string, blocking, or an error. How do these cases arise?
It could be the case that reading from \fd causes an error, e.g., it's
not an active file descriptor, points to a directory, etc.
It could be the case that \fd has enough text in the (symbolic) buffer
to read a whole line, either because an unescaped newline is actually
present or because there are no more writers to \fd, and so EOF has
been reached. In either of these cases, no scheduling need happen.
Finally, it could be the case that there isn't enough text to read a
line but there are still active writers. In this circumstance, a real
call to \texttt{read} would block. In \texttt{symbolic} mode, we
return the \pid of the first process that could write to \fd; that
process is stepped.
We must be careful, though: we don't want to block forever on the
writing process, only enough to finish reading the line. In this
circumstance, we allow the process to only step a few times before
checking the buffer again (hence the default of 10 steps).
\fi

Our scheduler is not a completely realistic model of POSIX
scheduling---we don't get to see every possible interleaving.
Our scheduling does, however, model the way pipes and processes
interact well enough for us to simulate interesting shell pipelines.
Our scheduler is acceptable in light of its benefits:
simple engineering;
API compatibility between \texttt{system} and \texttt{symbolic} modes;
a straightforward, linear visualization of shell stepping;
and determinism without rigidity.

\graf{Filesystem}

Our model of the filesystem is quite simplistic: we track a hierarchy of files and directories, but not file contents.
Such a simple model suffices to simulate pathname expansion and file
descriptor redirections ($\mathsf{dup}$) and heredoc redirections
($\mathsf{here}$), but not file redirections ($\mathsf{file}$).
It is a matter of engineering effort to produce a better symbolic filesystem.
It would be interesting to link our symbolic system up with
SibylFS~\cite{Ridge:2015:SFS:2815400.2815411} or
Forest~\cite{Fisher:2011:FLT:2034773.2034814}, or to use Ntzik et al.'s reasoning~\citeyearpar{Ntzik2017Reasoning,Ntzik2018Concurrent}.
We can also imagine implementing a read-only filesystem that allows
access to the real, underlying filesystem, but treats writes (and
other dangerous operations, like \texttt{execve}) as noops.
We consider platform-specific filesystems, like \texttt{/proc}, as out
of scope; we could in theory apply platform-dependent
reasoning~\cite{DBLP:conf/popl/NitaGC08}.

\section{POSIX conformance}
\label{sec:results}

Smoosh is meant to serve as a formal foundation for the POSIX shell.
But is Smoosh a good model?
We use three execution-based test suites to test Smoosh for conformance:
the official POSIX test suite, the Modernish test suite, and a suite of our own devising.
What are these test suites, why does using them support our claim, and how well do we conform?

\graf{The POSIX test suite}

The Open Group VSC test
suite\footnote{\url{https://www.opengroup.org/testing/testsuites/vscpcts2003.htm}}
is a broad set of tests for POSIX conformance systems.
We use VSC-PCTS2016 version 2.15.
\iffull
The test suite is built using TETware 3.8, a multi-language testing
harness capable of distributed testing. The actual POSIX shell tests
are written using the POSIX shell itself (which makes getting started
a bit of a challenge!); the shell is tested mostly within itself, but
also using external commands, particularly \texttt{awk}.
\fi
The POSIX shell test suite has 494 tests\iffull, each corresponding to
an assertion from the standard P1003.3 for testing conformance to
POSIX. This document existed as IEEE 2003.-1996 until it was withdrawn
in 2009.
Of the 494 tests,\else: \fi 1 is `not in use', 31 are `untestable' and are checked by
hand during certification,
and 44 are locale-dependent and marked `unresolved' or `unsupported'.
There are 418 locale-independent tests; these are the tests we use to
compare against other shells.

\graf{The Modernish test suite}

Modernish is a substantial library for the POSIX
shell~\cite{Modernish} aimed at simplifying the shell language.
\iffull
From
the project description:
\begin{quotation}
Modernish aims to provide a standard library that allows for writing
robust, portable, readable, and powerful programs for POSIX-based
shells and utilities. It combines little-used shell settings and a
modular library of functions to effectively build a new and improved
shell language dialect on top of existing shells.
\end{quotation}
\else
It uses existing shell features and compatibility testing to construct a new, shell-like language on top of an existing shell.
\fi
Modernish is implemented in 12k SLOC of shell scripts. It is very
portable. To achieve this portability, Modernish runs a series of
diagnostics against its host shell, detecting a variety of bugs and quirks.
\iffull
The first layer of tests to ensures that the host shell doesn't have
fatal, showstopping bugs--in which case Modernish can't run at
all. One such bug is \texttt{FTL\_NOARITH}, indicating incomplete
POSIX shell arithmetics support.
The second layer of tests collects the bugs, quirks, and capabilities
the shell supports.
For example, \texttt{BUG\_ARITHINIT} indicates that the shell aborts
when arithmetic expansion encounters null or unset variables (which is
a violation of the POSIX specification).
Quirks are implementation choices deemed to go against the grain (or,
at least, be in the minority among the shells Modernish
considers). For example, \texttt{QRK\_BCDANGER} indicates that inside
of a loop, an inner function call can use \texttt{break} and
\texttt{continue} to break out of the non-lexically enclosing loop.
Such a behavior is explicitly left unspecified in the specification.
Finally, capabilities are feature extensions. For example,
\texttt{ROFUNC} denotes the ability to set functions as read only.
\fi

\graf{Our test suite}

Our third test suite is our own. Our test suite has two parts:
internal unit tests and external system tests. Both parts serve as regression tests, but we can only use the external system tests on other shells, as the 
internal unit tests run Smoosh subsystems in symbolic mode.
\iffull
There are
573 unit tests broken into four sets tests: 253 tests of arithmetic,
27 tests of the filesystem and globbing, 64 tests for expansion (which
also tests how arithmetic and globbing work in the broader context of
expansion), and 229 for evaluation (which also tests how expansion
works in the broader context of evaluation).
\fi
Each external system test pairs a short shell scripts with expected
output and exit status. These are ``whole system tests'' but often
test obscure corners of the shell's behavior. A test harness runs the
scripts and checks outputs for a given shell executable.
There are 161 external system tests, in five categories: 2 speed
tests\iffull (to prevent performance regressions)\fi, 67 tests of
builtin commands, 2 parsing tests, 82 tests of shell semantics, and 8
tests of the \texttt{sh} executable's interface.
Some of the external system tests are adapted from bugs found by the
POSIX test suite and Modernish\iffull---we added them to our test suite
because it is faster than running the POSIX test suite and more
immediately informative than running Modernish's tests\fi.
Our test suite is occasionally too picky, demanding particular exit
status  when any non-zero exit status would be conformant.

\graf{Why use these tests?}
We use the POSIX test suite because it is the de facto standard of
what POSIX compliance is. It is imperfect (see
Section~\ref{sec:bugs}), but covers a great deal of ground.
It is also a significant stress test of the shell, comprising 29k SLOC
of tests in addition to 58k SLOC of shell code in the harness (which,
by default, runs in the shell under test).
The Modernish test suite condenses a great deal of knowledge and
experience with real shells into a very small package.
Modernish's shell scripts also rely on detailed characteristics of
shell behavior---during development, the Modernish test suite exercised several bugs in Smoosh's
semantics that the POSIX test suite did not detect\iffull (e.g., a
\texttt{case} command where no pattern matches should have an exit
status of 0)\fi.
Finally, our test suite not only tests against our own regressions from development,
but also highlights corner cases not covered in other test suites\iffull
(e.g., whether redirections are expanded before assignments)\fi.
\iffull
We are interested in adding other tests: from \OSH, from \ksh, and from
NetBSD, to just name three.
\fi

\graf{Conformance}
\begin{table}

\iffull
\begin{tabular}{@{}lrrrrrrrr@{}}
              & Smoosh  & \bash{}${}^*$ & \dash     & \zsh{}${}^\dagger$ & \OSH      & \mksh   & \ksh  & \yash{}${}^*$ \\
              & 0.1     & 4.4-12(1)     & 0.5.8-2.4 & 5.3.1-4+b2        & 0.6.pre21 & 54-2+b4 & 93u+  & 2.43-1      \\[1em]

\multicolumn{9}{@{}l@{}}{\textit{POSIX test suite (418 tests)}} \\ \hline
Failing tests & 0       & 4 (8)         & 20        & $\times$          & $\times$  & 35      & 23    & 22 (23)     \\
Time to run   & 12m41s  & 2m43s         & 2m43s     & $\times$          & $\times$  & 2m52s   & 3m24s & 2m45        \\

\fullbreak

\multicolumn{9}{@{}l@{}}{\textit{Modernish's shell diagnosis (91 potential bugs, 22 potential quirks) and test suite (312 tests)}} \\ \hline
Bugs          & 1       & 16            & 2         & 3                 & $\times$  & 3       & 14    & 1          \\
Quirks        & 0       & 4             & 2         & 5                 & $\times$  & 2       & 3     & 8          \\
Failing tests & 1       & 20            & 3         & 3                 & $\times$  & 3       & 17    & 1          \\
Time to run   & 5.5s    & 4.8s          & 1.4s      & 1.2s${}^\dagger$   & $\times$  & 3.2s    & 2.2s  & 2.4s        \\

\fullbreak

\multicolumn{9}{@{}l@{}}{\textit{Smoosh's test suite (161 tests)}} \\ \hline
Failing tests & 0       & 30 (35)       & 42        & 52                & 77        & 34      & 41    & 43 (39)     \\
Time to run   & 23s     & 28s           & 1m13s     & 42s               & 2m22s     & 21s     & 28s   & 29s         \\
\end{tabular}
\else
\begin{tabular}{@{}lrrrrrrr@{}}
              & Smoosh  & \bash{}${}^*$ & \dash     & \zsh{}${}^\dagger$  & \mksh   & \ksh  & \yash{}${}^*$ \\
              & 0.1     & 4.4-12(1)     & 0.5.8-2.4 & 5.3.1-4+b2         & 54-2+b4 & 93u+  & 2.43-1      \\

\multicolumn{8}{@{}l@{}}{\textit{POSIX test suite (418 tests)}} \\ \hline
Failing tests & 0       & 4 (8)         & 20        & $\times$           & 35      & 23    & 22 (23)     \\
Time to run   & 12m41s  & 2m43s         & 2m43s     & $\times$           & 2m52s   & 3m24s & 2m45        \\

\fullbreak

\multicolumn{8}{@{}l@{}}{\textit{Modernish's shell diagnosis (91 potential bugs, 22 potential quirks) and test suite (312 tests)}} \\ \hline
Bugs          & 1       & 16            & 2         & 3                  & 3       & 14    & 1          \\
Quirks        & 0       & 4             & 2         & 5                  & 2       & 3     & 8          \\
Failing tests & 1       & 20            & 3         & 3                  & 3       & 17    & 1          \\
Time to run   & 5.5s    & 4.8s          & 1.4s      & 1.2s${}^\dagger$    & 3.2s    & 2.2s  & 2.4s        \\

\fullbreak

\multicolumn{8}{@{}l@{}}{\textit{Smoosh's test suite (161 tests)}} \\ \hline
Failing tests & 0       & 30 (35)       & 42        & 52                 & 34      & 41    & 43 (39)     \\
Time to run   & 23s     & 28s           & 1m13s     & 42s                & 21s     & 28s   & 29s         \\
\end{tabular}
\fi

~ \\

\flushboth
\noindent
We use $\times$ to indicate that the tests could not be run\iffull (or ran but produced no output)\fi. \OSH's results are omitted (see Section~\ref{sec:bugs}).

\noindent
${}^*$Both \bash and \yash initiate a strict POSIX mode when run as
\texttt{/bin/sh}. The numbers in parentheses are the results from when
strict POSIX mode is turned off. Timings are from POSIX
mode. Modernish only uses POSIX mode.
${}^\dagger$\zsh was run only in \texttt{emulate sh} mode, which
Modernish uses as well. \zsh crashes in the Modernish test suite when
run noninteractively, so the timing is inaccurate.

  \caption{Comparison of shells on the POSIX test suite}
  \label{tab:tests}
\end{table}

We summarize the results of our various test suites in
Table~\ref{tab:tests}.
Of all of the shells tested, Smoosh is the only one to have no failing
tests in the POSIX test suite or in our own.
In Modernish, Smoosh has no quirks and one bug: Smoosh's parser (borrowed from \dash) cannot handle multibyte characters or characters with codes over 128; this bug triggers the one Modernish test that Smoosh fails.

\subsection{Bugs found}
\label{sec:bugs}

We found several POSIX compliance bugs in our two primary reference shells,
\dash and \yash, rediscovering a subtlety in the semantics for
\texttt{printf} that had already been independently
addressed.\iffull\footnote{\url{http://austingroupbugs.net/view.php?id=1205}}\fi{}
We also identified a typographical error in the POSIX spec and several
bugs in the POSIX test suite.
Identifying and reporting these bugs makes up only a small part of the
dividends of formal semantics, but we mention them to
highlight that even before we've significantly applied the
semantics, the process of development has been useful.

\graf{Bugs in shells}

In \dash, there were several issues: in arithmetic expansion,
variables that were unset or empty were improperly treated; the
\texttt{times} command reported incorrect numbers; and the
empty alias was mishandled. We submitted patches for these bugs; the first was
superseded by a different, independent fix of the same bug a year
later; the second and third are under review.
In \yash, asynchronous commands (e.g., \texttt{curl ... ~ \amp}) do
not have their STDIN redirected to \texttt{/dev/null} and \texttt{fg}
issues too much output.~\TODO{patches?}
Other shells clearly have significant bugs, but we have not had the
time to track them down in detail and report them.
Neither \zsh nor \OSH can run the POSIX test suite; \OSH cannot run the
Modernish suite, either, but it can run our suite, where it fails 77 of our tests.

\graf{Bugs in the POSIX test suite and specification}

\newif\iftp\tpfalse
\newcommand{\tp}[1]{\texttt{tp#1}}

We found ten bugs in the POSIX test suite, all of which have been
confirmed as true bugs and will be fixed in the next version of the
test suite.
\iffull
A test of numeric positional paramters \iftp(\tp{326})\fi incorrectly
uses the K shell syntax \texttt{function func\_sh5\_326 \{ ... \}} to
define a function, when the POSIX syntax is \texttt{func\_sh5\_326()
  \{ ... \}}.
A test of how the variable \texttt{\$\at} is treated \iftp(\tp{300})\fi had incorrect syntax for a \texttt{[} command.
A test of read-only variables used outdated semantics concerning
whether shells would exit when attempting to write to such variables
\iftp(\tp{422})\fi; finding this bug highlighted another, related issue
for the test suite developers\iftp (\tp{421})\fi.
A set of five tests concerning \texttt{break} and \texttt{continue}
misinterpreted the current POSIX standard, requiring that those
commands succeed when not in loop, when in fact such behavior is
unspecified\iftp (\tp{597}, \tp{600}, \tp{601},
\tp{607}, \tp{608})\fi.
A test of signal handling \iftp(\tp{718})\fi uses \texttt{kill -TERM \$\$} to kill the current process, but that syntax is an extension---it should instead use \texttt{kill -s TERM \$\$}.
Another test of signal handling \iftp(\tp{722})\fi executes undefined behavior when it sets a trap for \texttt{SIGKILL}.
\fi
We also found typographical errors in the POSIX specification and in
the POSIX test suite.
We also discovered a number of important shell behaviors that were
\emph{not} being tested\iffull; for example:
that redirections are expanded before assignments in a simple command
when a command is present and not a special builtin;
that redirections can have indirect targets;
exit statuses for various command\iffull (e.g., \texttt{case} commands that fall through and function definition)\fi;
that \texttt{for} loops respect read-only variables;
that the \texttt{command} builtin works on keywords;
and a variety of subtle tilde expansion behaviors
\fi. 
We are planning to submit our new tests to be added to the POSIX test suite.

\subsection{Performance}
\label{sec:testperf}

Smoosh is substantially slower than existing
implementations---about 4x slower on the POSIX test suite, slightly
better in Modernish (Table~\ref{tab:tests}).\footnote{Tests were run
  in Docker on an 2.8 GHz Intel Core i7 with 16GB RAM. Timings are
  from a single run, but there is little variance between runs.}
The timings in our own test suite should not be taken too seriously: the
only way to fail some of our tests is to time out.\iffull\footnote{For example, \dash
cannot complete our test suite without intervention due to incorrect
behavior when both \texttt{-i} and \texttt{-c} flags are provided and
an error occurs (\texttt{semantics.interactive.expansion.exit.test}).}\fi{}
Why is Smoosh slow?
Most shells are implemented as recursive evaluators in C, performing
expansion by mutation of compact data structures,
whereas Smoosh is implemented as iterated small-step semantics in OCaml (via Lem), performing expansion on immutable, non-compact ASTs.
Smoosh's slowness is not perceptible to us at the command line---most
interactive sessions spend the majority of their
wall clock time running executables, not scripts.

We speculated that some of Smoosh's slowness was because we weren't
hashing the filesystem calls in command name
\texttt{\textdollar{}PATH} resolution. We implemented
hashing, but our performance on the POSIX test suite was unchanged. Hashing may not be an optimization on modern systems\iffull; this question bears further study\fi.

\TODO{more detailed benchmarking?}

\section{Beyond the POSIX specification}
\label{sec:beyond}

The POSIX specification defines a common core for shells to implement,
but every shell has to make decisions about what is left unspecified. Every shell we encountered implements extensions, too.
In order for Smoosh's semantics to be a good model of the practical, implementation-oriented interpretation of the POSIX shell, we must understand how shells handle unspecified behavior, underspecified behavior, and extensions.
We offer examples of each below.

When evaluating \lambdajs, Guha et al.~ use the Mozilla test
suite---they ensure that they get the same answers as Rhino, V8, and
SpiderMonkey~\cite{10.1007/978-3-642-14107-2_7}.
Implementations of the POSIX shell do not have nearly the same level of
agreement as JavaScript interpreters do.
When developing Smoosh, we compared to many other shells, which
frequently disagree in corner cases. All of the shells implement
unspecified extensions of the POSIX standard, too.
Since none of these shells is perfectly POSIX compliant, we declined
to precisely match any of them. We strived instead for POSIX conformance,
clear semantics, and a lack of quirks.

\subsection{Unspecified behavior: non-local \texttt{break} and \texttt{continue}}
\label{sec:nonlocalcontrol}

Consider the case of non-lexical control. Does the following program print \texttt{hi} or not?
\begin{verbatim}
f() { break; echo hi; }; while true; do f; break; done  
\end{verbatim}
According to the POSIX specification, the \texttt{break} command in
the function \texttt{f} is not lexically enclosed in the loop. 
Non-lexical use of the control builtins \texttt{break} and
\texttt{continue} is left unspecified.
Shells behave differently! The more sensible option is to
forbid such non-lexical control, as most shells do, printing \texttt{hi} (and possibly a diagnostic message from the non-lexically enclosed \texttt{break} in \texttt{f}). But 
\bash 3.2.57(1) (which comes with OS X) and \zsh both allow the \texttt{break}
to pass through the function call and exit the non-lexically enclosing loop.
In Modernish, \zsh and the old \bash's behavior is called a \emph{quirk}\iffull, named
\texttt{QRK\_BCDANGER} because \texttt{break} and \texttt{continue}
are dangerous: user functions could break out of loops in the
Modernish library\fi.
There is also ambiguity around the permissible
behaviors of \texttt{break} and \texttt{continue} when used without a
non-lexically or with no enclosing loop at all\iffull: the standard says
that ``If there is no enclosing loop, the behavior [of \texttt{break}
  and \texttt{continue}] is unspecified.'' It would seem that shells
could then implement whatever behavior was desired, but the
specification also mandates that these commands \emph{must} exit with
0 unless a non-positive number is given as an argument; builtins can
only emit diagnostic messages on STDERR when the exit status is
non-zero, and so shells cannot warn via exit status or message that the
\texttt{break} or \texttt{continue} has no effect\else: the standard seems to forbid warning the user when control commands have no effect\fi.
The issue is under discussion with The Open Group.

\iffull
We must make some particular implementation choice. As a default, Smoosh
use lexical control, the safer option supported by \dash and \yash
where \texttt{break} and \texttt{continue} can't have effect through
function calls.
But a user can run \texttt{set -o nonlexicalctrl} to behave like the older \bash.
While it's impossible for Smoosh to `agree' with all shells, Smoosh is
at least able to express each of their behaviors.
In the extreme, we could offer flags for every quirk or implementation
choice, with special metaflags for selecting a subset (e.g.,
\texttt{set -o bashlike} would have Smoosh make unspecified decisions
following \bash).
It's not clear to us to what degree this is useful. Each such flag
adds cases to the semantics that will result in corresponding branches
in reasoning using the semantics.
A first step in determining whether such flags are useful would be to
examine a large corpus of shell scripts (as in the analysis of Debian
maintainer scripts~\cite{jeannerod:hal-01513750}) to determine which
unspecified behaviors they rely on.
\else
Prioritizing safety and avoiding quirks means that Smoosh uses lexical control (and will print \texttt{hi}), though non-lexical control can be enabled with a flag.
\fi

\subsection{Underspecified behavior: \texttt{getopts} and hidden state}
\label{sec:getopts}

The POSIX specification mandates that the \texttt{getopts} builtin
should use the user-visible shell variables \texttt{OPTIND} and
\texttt{OPTARG} to parse command-line and function arguments.
The rules are subtle, but the general idea is that \texttt{OPTIND} tracks the index of the current option argument; the \texttt{getopts} utility also takes the name of a variable to set with the current option. If \texttt{getopts} finds an option that takes an argument, the argument value is stored in \texttt{OPTARG}.
There's a problem in the POSIX specification, though: \texttt{getopts} needs more information than \texttt{OPTIND} and \texttt{OPTARG} to work properly. In particular, \texttt{getopts} needs to keep track of the offset into the current argument to handle `grouped' arguments.
\iffull
Why?
The POSIX specification says that \texttt{getopts} ``shall support
Utility Syntax Guidelines 3 to 10, inclusive''; in particular, Utility
Syntax Guideline 5 says, ``One or more options without
option-arguments, followed by at most one option that takes an
option-argument, should be accepted when grouped behind one '-'
delimiter''.
\else

\fi
As a concrete example, consider \texttt{getopts "ab:c:" opt -ab hi -c
  hello}, where \texttt{"ab:c:"} is the \emph{optstring} specifying
that \texttt{-a} is an option without arguments and that \texttt{-b}
and \texttt{-c} are options that take arguments, \texttt{opt} is the
variable name to set with the current option, and \texttt{-ab hi -c
  hello} is the argument list to process.
\iffull
\begin{sh}
getopts "ab:c:" opt -ab hi -c hello # OPTIND=1 or 2 opt=a OPTARG=      ?=0
getopts "ab:c:" opt -ab hi -c hello # OPTIND=3      opt=b OPTARG=hi    ?=0
getopts "ab:c:" opt -ab hi -c hello # OPTIND=5      opt=c OPTARG=hello ?=0
getopts "ab:c:" opt -ab hi -c hello # OPTIND=5      opt=? OPTARG=      ?=1
\end{sh}
Note that the `grouped' option \texttt{-ab} fits Guideline 5's rubric
of what can be grouped.
\fi
After parsing the first option, \texttt{opt} ought to be \texttt{a},
but what should \texttt{OPTIND} be? And where should the shell record the fact that the \texttt{a} has
already been processed?
\iffull
One possiblity would be to scan the first argument for an occurrence
of \texttt{a} and start after that, but that technique would not work
with repeated options, which are generally allowed but also
unspecified; see Guideline 11 and \S{}12.1 Utility Argument
Syntax. We haven't observed any shell with this implementations
strategy.
\fi
The choice we've seen taken in shells is to keep some extra
state to record the offset of the next character to process inside of a grouped option. Every
shell but \yash keeps this state hidden; Smoosh stores it in the
$\sigma.n^?_{\mathsf{optoff}}$ variable. \yash makes this state visible by
adding the current offset as in \texttt{OPTIND=1:2}.
Even so, shells differ slightly in their behavior\iffull.
Some shells (Smoosh, \dash, and \mksh) are early incrementers, setting \texttt{OPTIND} to 2 after processing \texttt{-a}.
Others (\bash, \zsh, and \ksh) are late incrementers, setting \texttt{OPTIND} to \texttt{1} after
processing \texttt{a}.
\else, with some incrementing \texttt{OPTIND} earlier than others.\fi

The issue has two root causes. First, the POSIX specification only
\emph{implies} \iffull but never overtly states \fi that more state is
needed. Second, the specification is silent on how to handle the
implicit state, leading to divergent behaviors.
\iffull
The specification says that ``the getopts utility shall place ... the
index of the next argument to be processed in the shell variable
OPTIND''---are we to interpret that as meaning that the
late-incrementing group has the correct behavior?
\fi
The issue is under discussion on the Open Group mailing list.\iffull\blindurl{https://www.mail-archive.com/austin-group-l@opengroup.org/msg04112.html}\fi

\subsection{Unspecified behavior and extensions: scope and the \texttt{local} builtin}
\label{sec:local}

\newcounter{COMMENTLOCAL}
\newcommand{\NT}[1]{\refstepcounter{COMMENTLOCAL}\label{comment:#1}(\arabic{COMMENTLOCAL})}
\newcommand{\RF}[1]{(\ref{comment:#1})}

\begin{table}

\begin{tabular}{@{}l@{}cccccccc@{}}
                            & Smoosh  & \bash       & \dash     & \zsh         & \OSH           & \mksh         & \ksh         & \yash   \\
                            & 0.1     & 4.4-12(1)   & 0.5.8-2.4 & 5.3.1-4+b2   & 0.6.pre21      & 54-2+b4       & 93u+         & 2.43-1 \\ \hline
nested scope \NT{nested}    & \YES    & \YES        & \YES       & \YES        & \YES           & \NO           & \NO          & \NO    \\
\fullbreak
\texttt{local} \NT{local}   & special & builtin     & special   & reserved     & builtin${}^+$  & special       & \NO          & \NO    \\
readonly \NT{readonly}      & error   & silent      & error     & override     & override       & override      &              &        \\
initial \NT{initial}        & unset   & unset       & unset     & null         & unset          & unset         &              &        \\
\texttt{-p} \NT{dashp}      & \YES    & \YES        & \NO       & $\sim$       & \NO            & $\sim$        &              &        \\
\end{tabular}

~ \\[.5em]

\flushboth
\noindent
\small
\RF{nested}~Do assignments before function calls have nested scope?
\RF{local}~What kind of command is \texttt{local}? \zsh identifies it as a reserved word; \OSH's \texttt{type} command calls it a builtin but it is in fact a syntactically restricted reserved word. Neither \ksh nor \yash support \texttt{local}.
\RF{readonly}~If \texttt{x} is declared \texttt{readonly}, can a local \texttt{x} be defined? \bash silently ignores the local definition; \zsh, \OSH, and \mksh allow for a local override.
\RF{initial}~What is the initial value of \texttt{x} after running \texttt{local x}?
\RF{dashp}~Both \texttt{readonly} and \texttt{export} will dump a list of variables when invoked with the argument \texttt{-p}\iffull (or with no arguments, though this is unspecified behavior)\fi; is there such a flag for \texttt{local}? \dash does not implement \texttt{-p}. \bash implements it, revealing that \texttt{local} is a macro for \texttt{declare}. Both \zsh and \mksh implement \texttt{local} as a macro for \texttt{typeset}, though they also show non-local variables.

\caption{The \texttt{local} builtin and nested scope in different shells}
\label{tab:local}
\end{table}

The POSIX shell has dynamic scope.
\iffull
It is challenging to build modular systems without some notion of
encapsulation, and lexical or at least nested scoping is a critical way to
achieve modularity in standard languages.
The shell tends to use subshells---i.e., forking---as its
encapsulation principle.
Even so, the POSIX specification requires some notion of nested
scopes.
\fi
Variable assignments on a command line, as in
\texttt{LD\_PRELOAD=... cmd bar baz} have three possible behaviors,
depending on the nature of \texttt{cmd} (per
Section~\ref{sec:evaluation}).
If \texttt{cmd} is a special builtin, then the assignments are
globally visible.
If \texttt{cmd} is a program or a non-special builtin, then the
assignments are visible only to that program or builtin.
Finally, if \texttt{cmd} is a function, then the assignments are
visible during the dynamic extent of the function, but it is
unspecified whether or not the assignments are globally visible
afterwards: scope may or may not be `nested'.
\iffull
The standard is careful to stipulate, however, that if \texttt{cmd} is
a non-special builtin that happens to be implemented as a function,
the variable assignments must remain local only to \texttt{cmd}.
That is to say, the implementation of shell builtins like
\texttt{getopts} needs to be done with some form of limited,
non-dynamic scoping.
\fi
Half of the shells we considered implement nested scope for
functions (Table~\ref{tab:local}).

Many shells implement a \texttt{local} builtin, which has syntax
analogous to \texttt{export} and \texttt{readonly} (Table~\ref{tab:local}).
That is, one can write \texttt{local x=5} inside of a function to
declare a variable \texttt{x} that will be (globally) bound to
\texttt{5} until the function returns, when \texttt{x} will revert to
whatever value it had before \texttt{local} was used.
Considering that the facilities to implement \texttt{local} line up
nearly exactly with those needed for nested scope, it is unsurprising
that of the four shells with nested scope, only \mksh doesn't have \texttt{local}.
\OSH implements nested scope; our tests, however, revealed a
bug in
\texttt{getopts}\iffull\blindurl{https://github.com/oilshell/oil/issues/335}\fi{}
along with a ``serious bug'' with
scoping.\iffull\blindurl{https://github.com/oilshell/oil/issues/329}\fi{}
These bugs will be fixed in \OSH's next release.

We implemented \texttt{local} in Smoosh, erring on the side of
featureful safety.
By analogy to \texttt{export} and \texttt{readonly}\iffull---both commands
that can set variables---\else, \fi{}we've treated \texttt{local} as a special
builtin\iffull (Figure~\ref{tab:local}\RF{local})\fi. Since erroneous conditions
in special builtins cause shell scripts to abort, making this choice
assures early failure in case \texttt{local} is misused.
Like in \dash, it is an error in Smoosh to create a local variable
with the same name as a readonly variable. It may seem unduly
restrictive, since the variable will be locally scoped, but triggering
an error makes readonly variables unforgeable, even locally.
\iffull
We support the \texttt{-p} flag for printing out current local
variables.
\fi

\iffull

\section{Discussion}
\label{sec:discussion}

Smoosh is a \emph{particular} instance of the POSIX specification, not completely conforming to \dash, \bash, \yash, or any other existing implementation.
Nevertheless, we believe Smoosh is a useful interpretation of the POSIX specification.
Smoosh is, to our knowledge, the first implementation of the POSIX
shell to treat the shell state immutably/functionally. While less
efficient, the immutable shell state and small-step semantics combine
to yield a mechanized system for explaining how Smoosh evaluates.
We believe that some of the negative attitude towards the shell is
rooted in misunderstanding, and Smoosh will perhaps help researchers
better understand the shell.
Furthermore, the testing in Section~\ref{sec:results} establishes that
Smoosh is a conformant, quirk-free interpretation of the POSIX
specification.
We can make an analogy: the CompCert compiler doesn't exactly match
any particular existing compiler, but its semantics of C can still be
treated as `canonical'~\cite{Blazy2009,10.1007/978-3-319-08970-6_36}.
Our analogy will be made stronger as we use the Smoosh semantics to
build tools like, say, a verified compiler.

\graf{Interactivity}

While Smoosh is usable as an interactive shell, it lacks many of the
features that make \bash, \zsh, and Fish popular:
\texttt{readline}/\texttt{editline}-style command-line editing (e.g.,
with Emacs or Vi commands), interactive history, and command
completion.
We are very interested in the interactive features in shells, but
before we implement these features in Smoosh we plan to first collect
and taxonomize existing interactive shell features, and then separate
parsing and interaction in Smoosh.

\graf{Performance}

Smoosh is about 4x slower than the shells we tested against
(Section~\ref{sec:testperf} and Table~\ref{tab:tests}).
A recursive interpreter or a compiler would both speed up Smoosh
substantially. It is a natural plan to first write an interpreter (and
prove it is identical to the small-step semantics) and then to write a
compiler that corresponds to the interpreter.
There are many opportunities for optimization in a compiler, but we
must be careful: the shell offers many observations, and preserving
behavior exactly is a challenge.
For example, one can observe the number of processes allocated by
checking \pid{}s, so an optimization that avoided a subshell would lead
to slightly different behavior.
It may be prudent to settle for some kind of up-to equivalence.

\graf{Safety}

ShellCheck is a popular linter for the shell~\cite{ShellCheck}, working on syntactic principles.
We are interested in extending ShellCheck to be able to use Smoosh's
semantics for local reasoning.
More ambitiously, we could use our symbolic semantics (extended to support symbolic branching) to generate summaries of the actions taken by, e.g., installation scripts.
More intensive uses of the semantics might generate weakest
preconditions for a script to generate a zero exit status.
One might want a `hardening' compiler to then check those
preconditions on entry. Doing so is not, of course, behavior
preserving---it eliminates bad behaviors!

\graf{Design}

The POSIX specification is a living standard; can we use Smoosh's
semantics to help with the ongoing design of the POSIX shell?
There are several possible ends for design:
for safety, for intelligibility, for performance, or for new features.
Design discussions of the POSIX shell are generally held by
implementors (on, e.g., the Austin Group mailing list) without
reference to any body of scripts of particular interest.
The design process would benefit from collecting and analyzing shell
scripts~\cite{jeannerod:hal-01513750} to determine which features are
used, what unspecified or undefined behavior is executed, etc.
More sweeping changes are of course possible. For example,
user-defined parameter formatters would allow users to more easily
process strings.\footnote{At the risk of ``brain damage'',
  per~\url{http://zsh.sourceforge.net/Doc/Release/Expansion.html\#Parameter-Expansion-Flags}.}
Our experience with the memoization of command resolution results
(a/k/a hashing, Section~\ref{sec:testperf}) indicates that it may not
be a performance win on modern systems.
Modernish and \zsh both rethink field splitting and globbing; an
empirical analysis would allow us to determine how these things can be
relaxed.

\fi

\section{Related work}
\label{sec:relwork}

\graf{Research on the shell}

The POSIX shell has seen relatively little academic attention.
There are only two recent works that take the shell's semantics seriously: ABash~\cite{Mazurak07abash} and
CoLiS~\cite{jeannerod:hal-01513750,regisgianas:hal-01890044,jeannerod:hal-01432034,jeannerod:hal-01534747}.
ABash is a static analysis for \bash scripts in particular; it checks
for common expansion bugs along with taint tracking.
ABash has some limitations in its parser and in its model of
expansion; we hope that Smoosh's semantics can combine with ABash's
approach to provide a more precise analysis that can cover more of the
shell.
The CoLiS project takes a core calculus approach to the shell, not
unlike \lambdajs~\cite{10.1007/978-3-642-14107-2_7}. Jeannerod et
al. define an interpreter for a tiny shell-like
language~\citeyearpar{jeannerod:hal-01432034,jeannerod:hal-01534747} to which
they elaborate shell using their hand-built parser,
Morbig~\cite{regisgianas:hal-01890044}. The CoLiS interpreter is not
yet a usable shell---it passed only 8 of our 161 tests, mostly due to a variety of unsupported shell features (e.g., assignments in commands, heredocs, \texttt{break}); even so, its symbolic
evaluator has already found numerous bugs in Debian maintainer
scripts~\cite{jeannerod:hal-01513750}.
Smoosh has taken the ``full semantics'' approach rather than the
``elaborate to a core calculus'' approach; by implementing the full
semantics, we have a baseline against which we can prove an
elaboration correct. We believe it would have been \emph{much} more
challenging for us to achieve Smoosh's level of conformance if we had started with elaboration (see ``Whole-language semantics'' below).
We have left parsing out of our scope for now (see
Section~\ref{sec:limitations}); we hope to extend Smoosh to use Morbig
and other parsers.
Greenberg has made arguments about why certain features of the shell
are useful for concurrency~\citeyearpar{Greenberg18dsl} and
interactivity~\citeyearpar{Greenberg18px}; see the latter paper for
references on much older research on command-line interfaces,
interactivity, and live programming~\cite{Collins2003live}.
NoFAQ~\cite{DAntoni16NoFAQ} uses machine learning to suggest
repairs to commands, but treats shell syntax as unstructured text.

\graf{Tools for the shell}

Outside of academia, programmers have created a variety of tools to
support shell programming.
Modernish is a library that rebuilds the shell language using its own
features~\cite{Modernish}.
ShellCheck~\cite{ShellCheck} is a linter that can process a variety of shell extensions;
it is purely syntactic, though, and could be improved by having it
reason using Smoosh's semantics.
ExplainShell~\cite{explainshell} patches together a parser for
bash\footnote{\url{https://github.com/idank/bashlex}} and some
post-processed man pages to explain each part of a shell structure.
ExplainShell isn't about semantic insight: e.g., asking it about
\verb|${#*}| doesn't mention that its expansion is unspecified; nothing indicates the difference between \verb|$@| and \verb|"$@"|.
Finally, \texttt{maybe} allows for tentative use of shell
commands~\cite{maybe}.

\graf{Replacement shells, scripting languages, and shell libraries}

A variety of replacement shells have been proposed, both in academia,
with projects like Scsh, Shill, and
Rash~\cite{Shivers06scsh,MooreDKC2014,DBLP:conf/gpce/HatchF18} and
outside of academia, with popular interactive shells like fish and
\zsh, along with less well known replacement shells like Xonsh and
\OSH~\cite{Xonsh,Oil}.
Only \zsh and \OSH aim for any real POSIX compatibility; Xonsh strives
to have some measure of compatibility with \bash.
Others are more scripting languages than shells: Shivers planned but
never developed an interactive mode for Scsh; Shill isn't meant to be
interactive, either.
Shell libraries, like Plumbum, Turtle, and Shcaml aim to bring
shell-like idioms into conventional programming
languages~\cite{Plumbum,Turtle,DBLP:conf/ml/HellerT08}.
None of these quite match our scope: POSIX shells that can
be used both interactively and programmatically.

\graf{Whole-language semantics}

There have been a variety of efforts at building language semantics
for whole, real languages: for
JavaScript~\cite{Maffeis:2008:OSJ:1485346.1485368,10.1007/978-3-642-14107-2_7},
for
C~\cite{Blazy2009,Ellison:2012:EFS:2103656.2103719,10.1007/978-3-319-08970-6_36,Kang:2015:FCM:2737924.2738005,Memarian:2016:DCE:2908080.2908081},
and for
Rust~\cite{DBLP:journals/corr/abs-1806-02693,Jung:2017:RSF:3177123.3158154,DBLP:journals/corr/abs-1903-00982},
to name a few.
Approaches to such whole-language semantics fall broadly into two
styles of modeling: large semantics that cover a broad range of
language constructs, using light syntactic sugar; and small semantics
that cover a narrow range of language constructs, using heavy
syntactic sugar/elaboration.
K and Lem epitomize the former
style~\cite{rosu-serbanuta-2010-jlap,Mulligan:2014:LRE:2628136.2628143},
while \lambdajs is emblematic of the latter style, defining a small
calculus to which it elaborates
JavaScript~\cite{10.1007/978-3-642-14107-2_7}.
Why did we choose the large semantics route?
First, the shell's complexity makes it hard to precisely encode every
aspect of a shell construct in terms of other, simpler ones; moreover,
using a large AST lets us better match parts of the semantics to the
POSIX spec.
Second, \lambdajs leaves out \texttt{eval}---to include it requires
including not only a parser in the desugarer or semantics, but also
the desugarer itself.
We cannot do without \texttt{eval}~\cite{10.1007/978-3-642-22655-7_4}:
it's used in a variety of testing frameworks and is the only way to
achieve certain forms of indirection in the shell (e.g., a tilde
expansion with a variable).
Now that we have much more experience with the shell---and a reference
semantics against which to prove our elaboration correct---we can
think about desugaring and compilation\iffull (see
Section~\ref{sec:discussion}, ``Performance'')\fi.

\TODO{Cf. compcert's CLight semantics \url{https://github.com/AbsInt/CompCert/blob/master/cfrontend/Cexec.v}}
\TODO{how many `rules' would \texttt{step\_expansion} correspond to? what about \texttt{step\_eval}?}

\section{Conclusion}
\label{sec:conclusion}

Smoosh is a small-step operational semantics for a new shell
implementing the POSIX shell standard.
The executable semantics is implemented in Lem code but corresponds
well to legible, practicably complex inference rules.
Of the shells we have considered, Smoosh best conforms to the POSIX
specification and has the fewest bugs and quirks.
In the process of developing and testing Smoosh, we found numerous
bugs, confusions, and underspecifications in existing shells used in
production, in the POSIX test suite, and in the POSIX specification
itself.

Smoosh cleanly separates OS functionality from the core shell
semantics, allowing us to use our tested semantics in symbolic
settings; we demonstrate this capability with a stepper.
The semantics we've developed here promises to support research on and
development of shells and tools for shells. Our semantics will
also form the basis of design innovations and revisions in the shell
and, we hope, interactive programming in general.

\begin{acks}
  Much of this work was done while the first author was on sabbatical
  at Harvard University; he gratefully acknowledges the Harvard CS
  department in general and Steve Chong in particular for their
  support.
  Brian Selves of the Open Group provided invaluable help with
  understanding the POSIX specification and the POSIX test suite, the
  latter of which the Open Group helpfully provided.
  The Austin Group mailing list offered helpful, timely feedback
  and discussion.
  Discussions with Ralf Treinen, Yann R{\'e}gis-Gianas, and Andy Chu
  were helpful and interesting.
  Aaron Bembenek and David Holland had helpful comments.
  Hannah de Keijzer proofread.
\end{acks}

\bibliography{mgree}

%

\end{document}